\documentclass[final,3p,times,twocolumn,authoryear]{elsarticle}
\usepackage[colorlinks=true]{hyperref}
\usepackage{tabularx}
\usepackage{ragged2e}

\usepackage{amssymb}
\usepackage{amsmath}

\journal{New Astronomy Reviews}

\newcommand{\vect}[1]{{\mathbf{#1}}}
\newcommand{\rd}{{\rm d}}

\begin{document}

\begin{frontmatter}



\title{Radio galaxies and feedback from AGN jets}


\author{M.J. Hardcastle$^{1}$ and J.H. Croston$^{2}$}

\address{$^{1}$Centre for Astrophysics Research, University of Hertfordshire, College Lane, Hatfield AL10 9AB\\ $^{2}$School of Physical Sciences, The Open University, Walton Hall, Milton Keynes, MK7 6AA, UK}

\begin{abstract}
We review current understanding of the population of radio galaxies
and radio-loud quasars from an observational perspective, focusing
on their large-scale structures and dynamics. We discuss the physical conditions in radio galaxies, their fuelling and accretion modes, host galaxies and large-scale environments, and the role(s) they play as engines of feedback in the process of galaxy evolution. Finally we briefly summarise other astrophysical uses of radio galaxy populations, including the study of cosmic magnetism and cosmological applications, and discuss future prospects for advancing our understanding of the physics and feedback behaviour of radio galaxies.
\end{abstract}


\end{frontmatter}

\section{Introduction}

\label{sec:intro}
Radio galaxies and radio-loud quasars (collectively radio-loud AGN, or
RLAGN in this article) are active galaxies characterized by radio
emission driven by jets on scales from pc to Mpc. The characteristic radio emission
is synchrotron emission: that is, it indicates the presence of
magnetic fields and highly relativistic electrons and/or positrons.
Synchrotron emission may be seen in other wavebands, and this enabled
the detection of the first radio galaxy jet before the advent of radio
astronomy \citep{Curtis18} but it was only with the capabilities of radio
interferometry \citep{Ryle52} that it became possible to detect and image these
objects in detail and in large numbers. As we will discuss in more
detail below, radio observations remain key to an understanding of
their origin, dynamics and energetics.

\begin{figure*}
  \begin{center}
  \hskip 0.001pt
  \hbox{
    \includegraphics[width=0.40\linewidth]{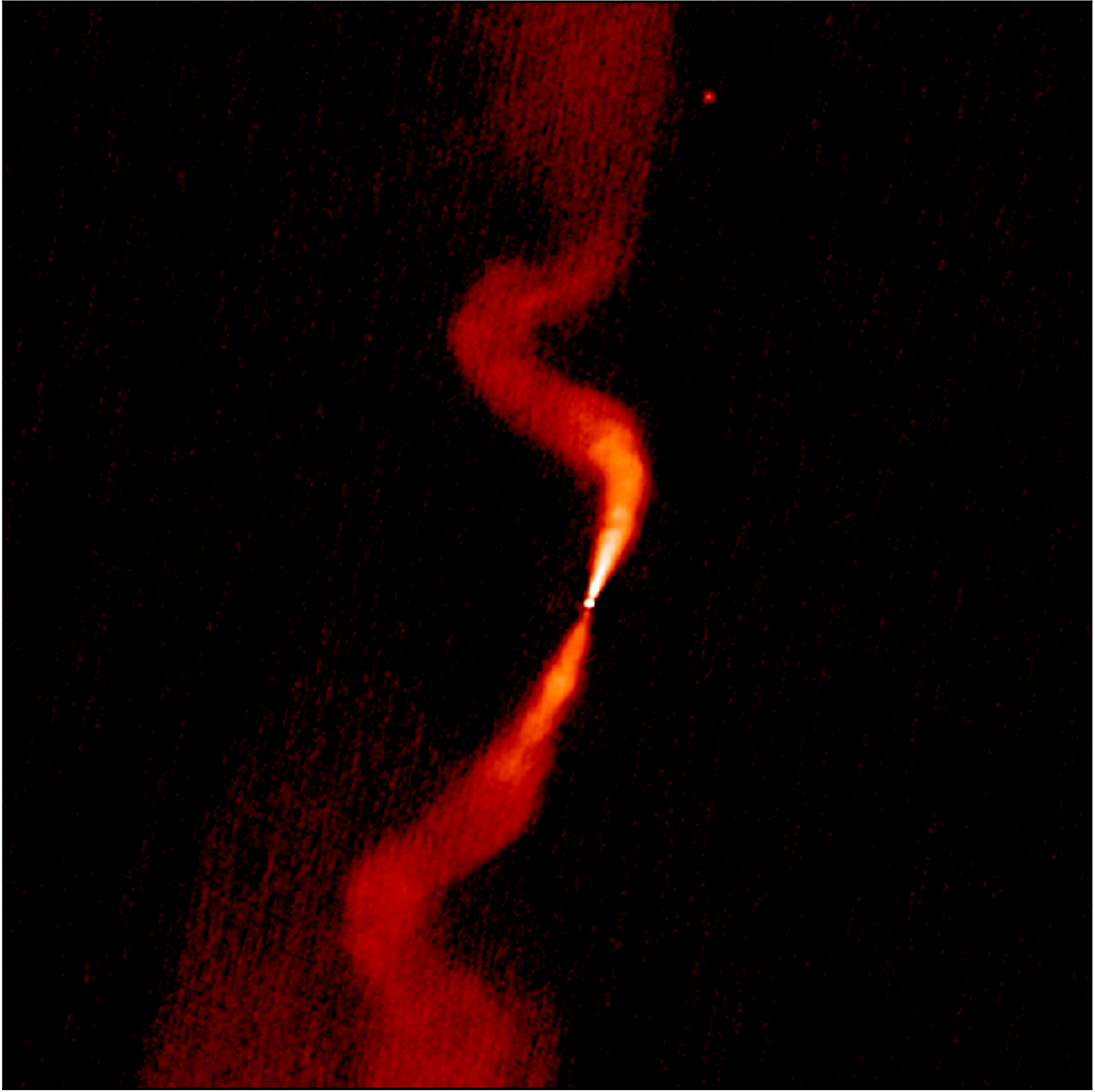}
    \includegraphics[width=0.40\linewidth]{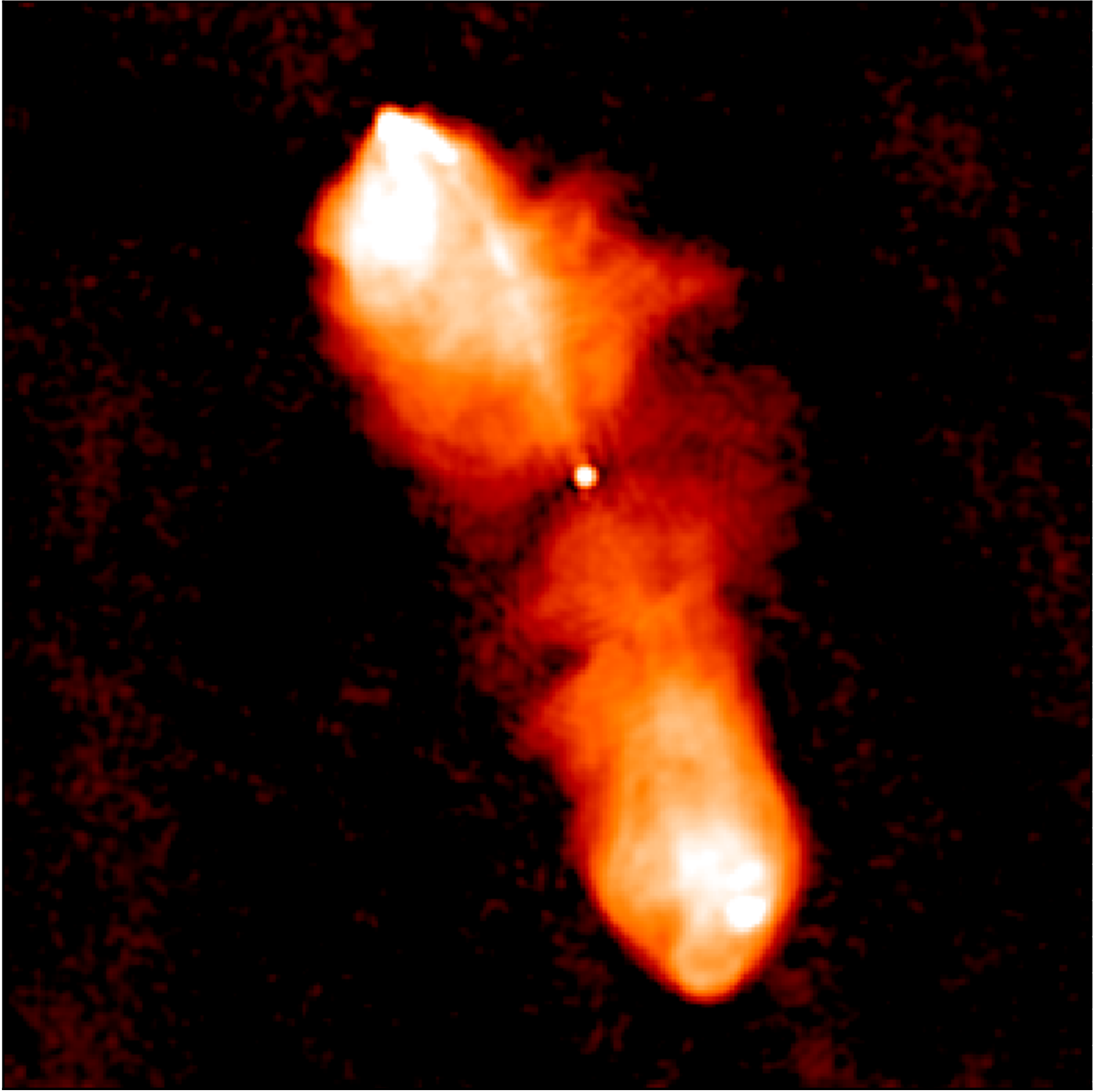}
  }
  \hbox{
  \includegraphics[width=0.40\linewidth]{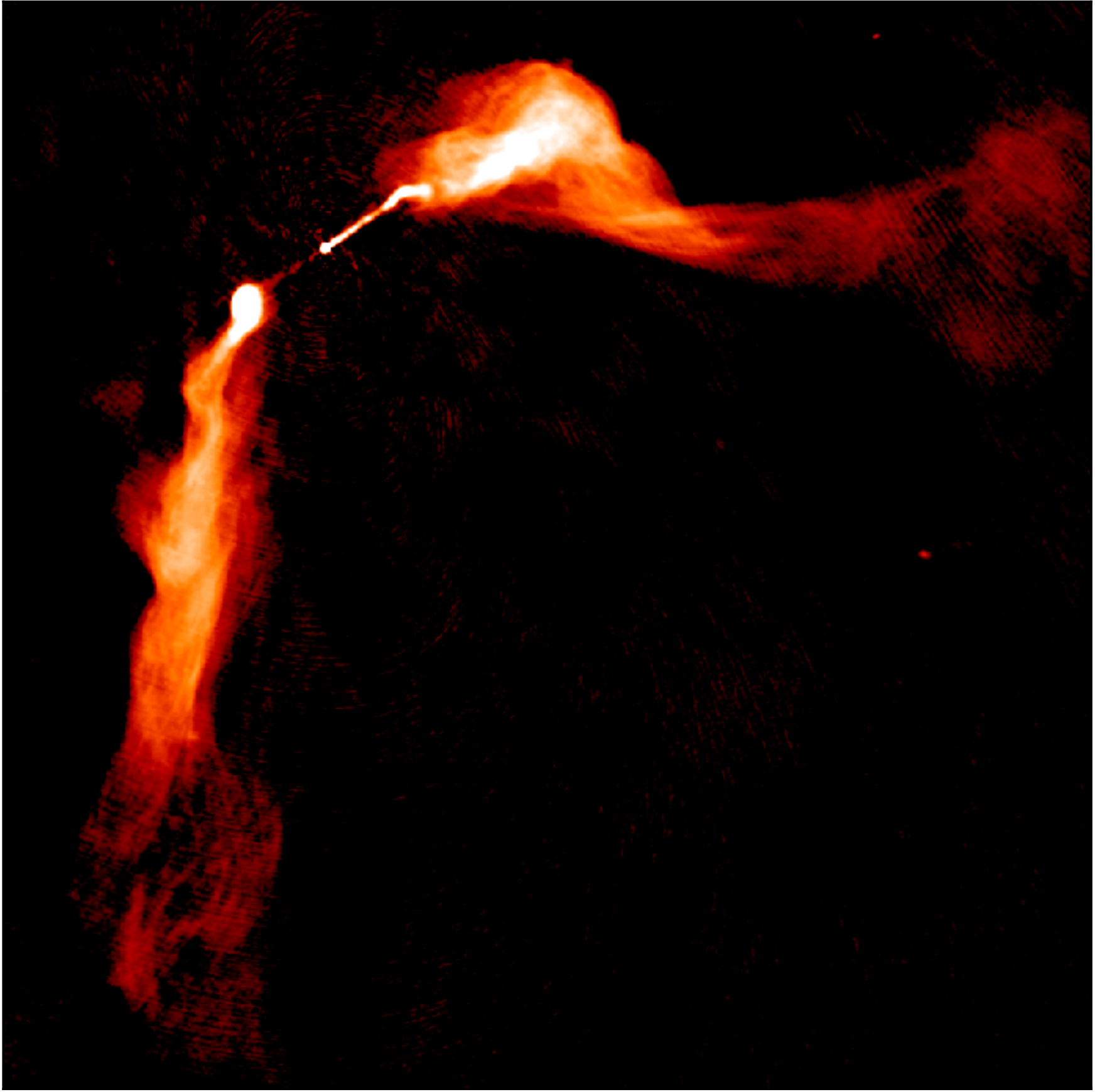}
  \includegraphics[width=0.40\linewidth]{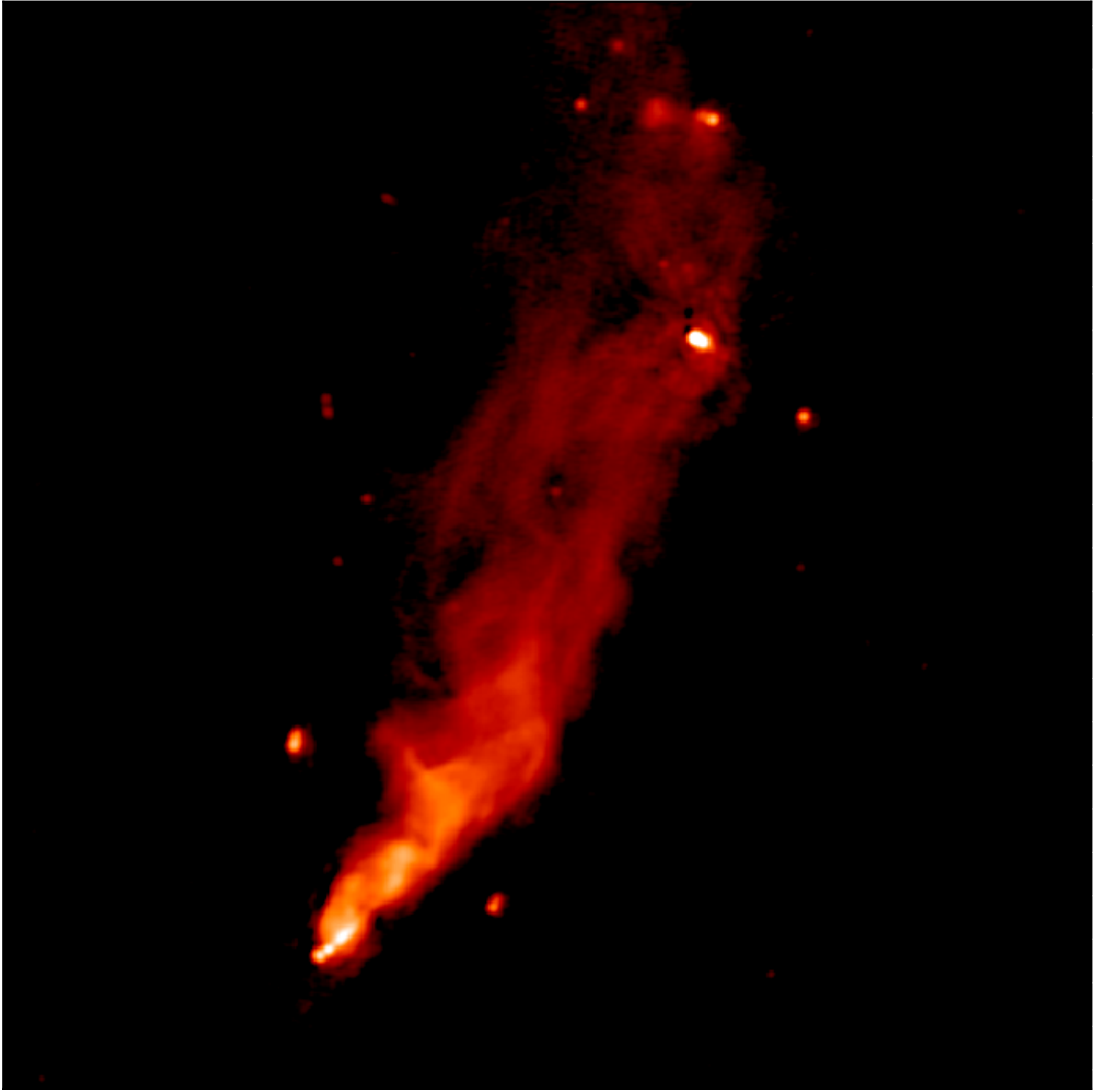}
  }
  \hbox{
  \includegraphics[width=0.40\linewidth]{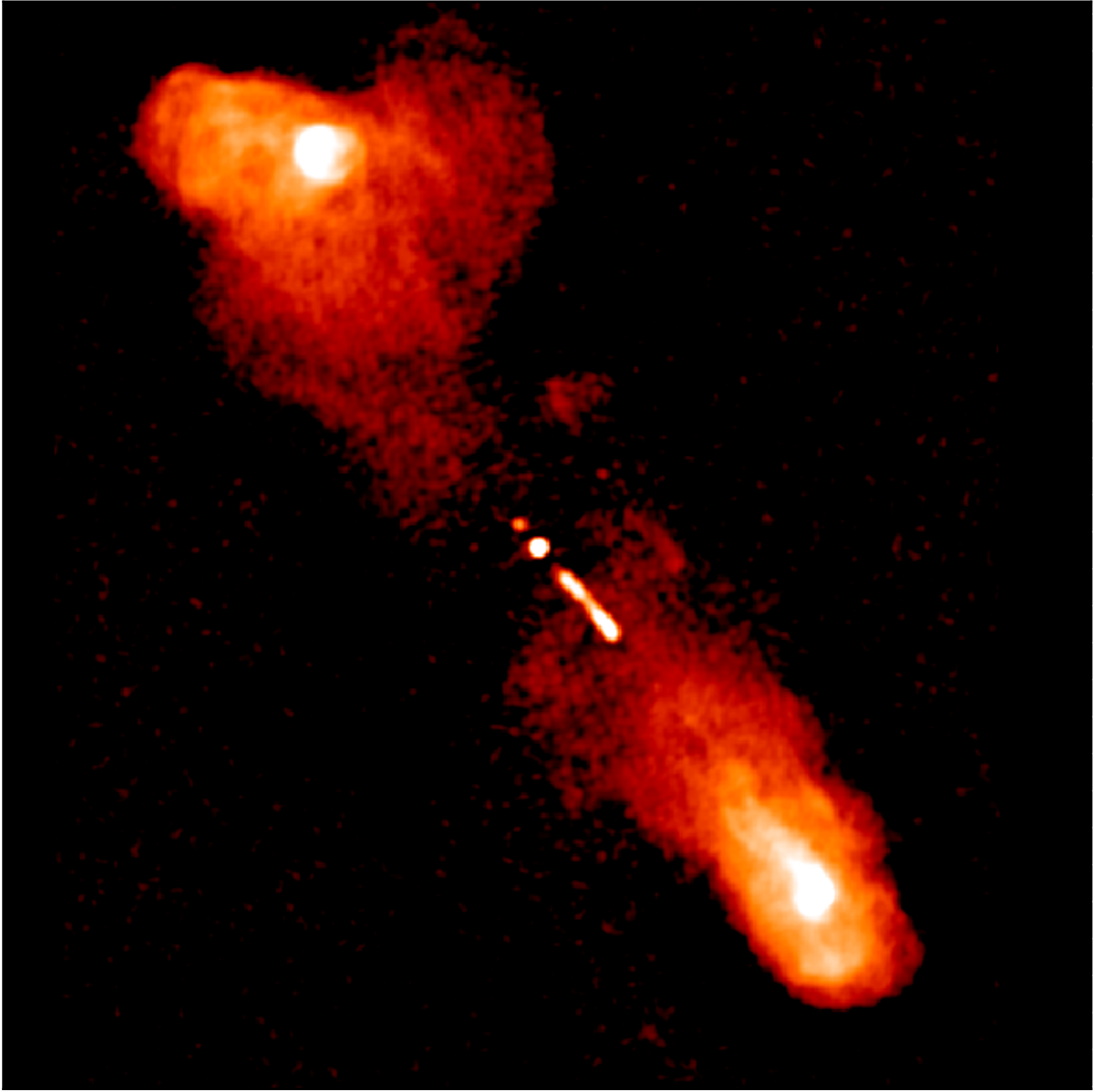}
  \includegraphics[width=0.40\linewidth]{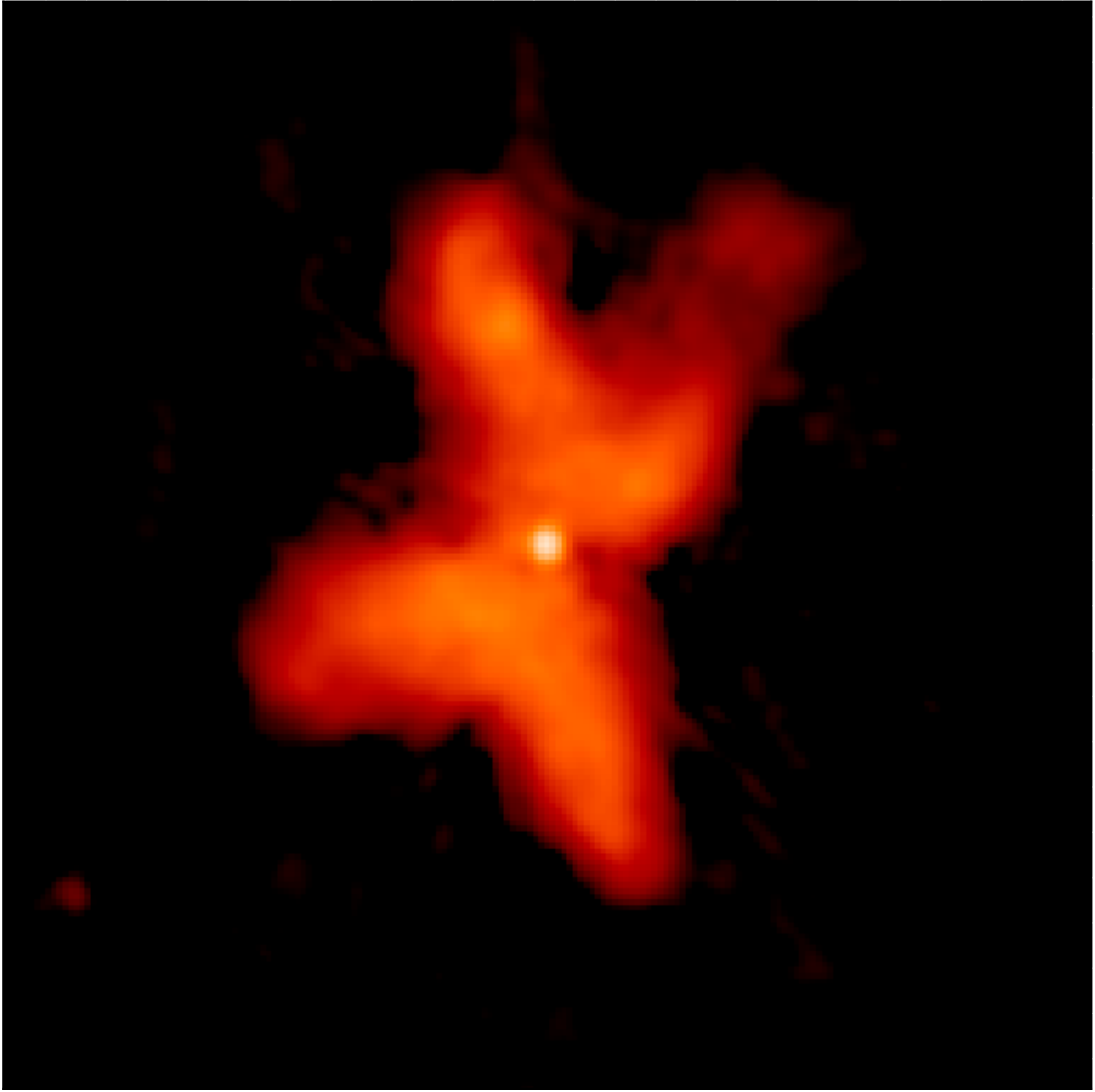}
  }
  \end{center}
  \vskip -16pt
  \caption{Radio images of nearby radio galaxies showing a range of
    morphologies: top row are the Fanaroff-Riley class I source 3C\,31
    (left) and the Fanaroff-Riley class II source 3C\,98; middle row
    are the wide-angle tail source 3C\,465 (left) and narrow-angle
    tail / head-tail source NGC\,6109 (right); and bottom row are
    double-double radio galaxy 3C\,219 (left), and core-restarting
    radio galaxy 3C\,315 (right). Compact `cores' may be seen in all
    images, well-collimated jets are visible in 3C\,31, 3C\,98 and
    3C\,465, and hotspots in 3C\,98, 3C\,465 and 3C\,219. 3C\,31 image kindly provided by
    Robert Laing; 3C\,98 image from the online `Atlas of DRAGNS' at
    \url{http://www.jb.man.ac.uk/atlas/}; 3C\,465 image courtesy of
    Emmanuel Bempong-Manful; 3C\,219 image from \cite{Clarke+92}; NGC
    6109 and 3C\,315 from unpublished LOFAR data.}
  \label{fig:agn-picture}
\end{figure*}

Radio images of some characteristic large-angular-scale
nearby RLAGN are shown in Fig.\ \ref{fig:agn-picture}. These show the
large-scale jets and lobes that are the defining feature of this type
of object. The first observations capable of showing the radio jets
\citep[e.g.,][]{Northover73} motivated the development of the now
standard `beam model', in which collimated outflows from the active
nucleus drive the extended structures
\citep{Longair+73,Scheuer74,Blandford+Rees74}. Some authors have used `beam' to
refer to the outflows themselves and `jet' to refer to their
observational manifestations, but in this review we use `jet'
interchangeably for both, relying on context to make the distinction
clear where it is needed.

Key historical developments in observational RLAGN studies after the
first surveys and optical identifications included the development in
the 1970s of high-resolution interferometers such as the 5-km
telescope and the NRAO Very Large Array (VLA), which allowed detailed
study of radio structures as well as optical identifications for the
first time; progress in very long baseline interferometry (VLBI),
which has given increasingly detailed views of the inner parts of the
jets; the advent of sensitive optical telescopes, including the {\it Hubble
Space Telescope}, which allowed detailed studies of RLAGN host galaxies
and environments in the optical, as well as the study of optical
synchrotron radiation; a greatly improved understanding of the nature
of the active nuclei themselves, driven by a combination of broad-band
photometry and spectroscopy; and the development of X-ray telescopes
with the sensitivity needed to image both the hot gas environments of
RLAGN and the X-ray synchrotron and inverse-Compton emission from the
large-scale radio structures. Some of the understanding derived from
those observational advances is discussed in later sections of this
review. However, perhaps the most important development has been the
realization that the energetic input of RLAGN can have a profound
effect on both the galaxies that they inhabit and their large-scale
environment, heating the hot gas that surrounds them and preventing it
from cooling and forming stars; this process is an important member of
a family of processes that have come to be called `AGN feedback'. This
understanding of the importance of RLAGN in galaxy formation and
evolution, derived both from X-ray observations and from numerical
modelling of the formation and evolution of galaxies, has moved RLAGN
studies into the mainstream of extragalactic astrophysics. In this
chapter we will therefore also discuss how observations and models of
RLAGN constrain the `feedback' processes that may be operating.

Throughout the review we adopt the convention that $\gamma$ represents
the (random) Lorentz factor of an individual electron and $\Gamma$
represents the bulk Lorentz factor due to directed motion.
Luminosities and physical sizes quoted are based on a standard concordance cosmology with $H_0 = 70$ km s$^{-1}$ Mpc$^{-1}$.

\section{Observational approaches}

\label{sec:obs}
In this section we provide an overview of the observational methods that provide us with constraints on radio galaxy physics.

\subsection{Radio}

The fact that the radio emission is synchrotron emission was realised
early on from its polarization and spectrum \citep{Baade56,Burbidge56}
and, together with the optical identification of these objects with
relatively distant galaxies (see Section \ref{sec:opticalid}), turned
out to imply very large energies stored in the extended structures.
The details of the radiation mechanisms can be found in e.g.,
\cite{Longair10} or \cite{Rybicki+Lightman79}. The point that we wish
to emphasise here is that
the energy density in the radiating electrons and field, $U$, can be
written in terms of the volume emissivity $J(\nu)$ and the magnetic
field strength $B$: for a power-law
distribution of electron energies with energy index $p$, we find
\begin{equation}
  U = k J(\nu) B^{-\frac{p+1}{2}} + \frac{B^2}{2\mu_0}
  \label{eq:u}
\end{equation}
where $k$ is a constant incorporating physical constants, the
observing frequency, and the integral over electron energies. Clearly
eq.\ \ref{eq:u} has a {\it minimum} at some value of $B$, and by
solving for the minimum and computing the minimum energy density, we
can get both an estimate of a characteristic field strength and a
lower limit on the energy responsible for a given region of a radio
source. The minimum-energy condition turns out to be close to the {\it
  equipartition} energy, $U_e = U_B$, but the important conclusion is that the
minimum-energy field strengths for 100-kpc-scale lobes such as those
shown in Fig.\ \ref{fig:agn-picture} are of order 1 nT for a powerful
source, leading to energy densities of order a few $\times 10^{-13}$ J
m$^{-1}$ and total energies of order $10^{54}$ J or more --- which
would require the direct conversion to energy of millions of solar
masses of matter. If there is any departure from the minimum-energy
assumptions, these numbers will be larger --- and possibly very much
larger if, e.g., there are large departures from equipartition or if
the energy density in the lobes is dominated by non-radiating
particles.

It can be seen that estimates of the energetics of the radio-emitting
structures depend strongly on the characteristic magnetic field
strength. This cannot be estimated directly from observations of
synchrotron intensity. Synchrotron emission is strongly polarized ---
the fractional polarization can be $\sim 70$\% for a uniform-field
region with a power-law spectrum, and even higher where the spectrum
is exponentially cutting off --- but the polarization does not tell us
about the field strength either, although it does give an
emission-weighted estimate of the magnetic field {\it direction} along
a particular line of sight, if Faraday rotation effects may be
neglected (see below). For optically thin radio emission, the only way
of directly estimating the magnetic field strength is to use
additional observations, for example observations of inverse-Compton
emission, discussed below (Section \ref{sec:x-ray}).

Faraday rotation is an effect caused by the propagation of
electromagnetic radiation through a magnetised, ionized medium. The
polarization angle rotates due to a difference in propagation speed
for the two circularly polarized components of the electromagnetic
wave. The change in angle is dependent on frequency, and the rotation
measure --- the strength of the rotation effect --- depends on the
magnetic field strength and electron density of the intervening
material \citep[e.g.][]{Cioffi+Jones80}. For a single line of sight
through a Faraday-active medium towards a background polarized source
the measured polarization angle $\chi$ is given by \citep{Burn66}:
\begin{equation}
  \chi = \chi_0 + \phi \lambda^2
\end{equation}
where $\chi_0$ is the intrinsic polarization angle and $\phi$ is given by
\begin{equation}
  \phi = K \int_0^d n_e \vect{B} \cdot \rd \vect{S}
\end{equation}
in which $K$ is a constant with value (in SI units) $2.63 \times
10^{-13}$ T$^{-1}$. Clearly in general different lines of sight, even
within a given telescope beam, will
have different values of $\phi$. However, observationally, it is often
the case that the rotation measure $RM = \rd \chi / \rd \lambda^2$
shows smooth behaviour across a source, and in this case rotation
measure observations can be used to estimate magnetic field strength
{\it external} to the source
along the line of sight in
situations where the density of intervening plasma is known or can be
estimated. Frequency-dependent depolarization also provides
information about the magnetic field strength and/or the density of
thermal plasma internal and external to the radio lobes
\citep[e.g.][]{Burn66,Laing88}, although disentengling the
contributions of different components can be challenging.

Although RLAGN are now detected at many other wavebands, radio
observations continue to provide the most efficient method of {\it
  selecting} them. Early surveys such as the 3C or Parkes surveys
\citep{Bennett62,Bolton+64}, and the optically identified catalogues
derived from them \citep[e.g., 3CRR,][]{Laing+83}, have been the
background for many of the detailed studies of the physics of
individual AGN or small samples. Because RLAGN are a comparatively
rare population, wide sky areas are necessary to get a representative
view of the local population, and so historically low-frequency radio
telescopes, with their large fields of view but comparatively low
sensitivity, were the survey instruments of choice. Low-frequency
selection has the advantage that the low-frequency emission of RLAGN is
dominated by the steep-spectrum emission from the lobes, presumed to
be more or less isotropic. More recently surveys at GHz frequencies, especially with the
Very Large Array (VLA), have been used to generate large samples.
Wide-area surveys can be used to find many objects in the local
universe \citep{Best+Heckman12} while deep surveys with a narrow field
of view probe the RLAGN population at high redshift
\citep[e.g.,][]{Smolcic+17}. In these sensitive surveys, separating
the radio emission from RLAGN from that due to star formation in the
host galaxy becomes a major concern. Forthcoming wide and/or deep
surveys with LOFAR \citep{Shimwell+17}, ASKAP \citep{Norris+11}, and
MeerKAT \citep{Jarvis+17} will provide still larger samples of the AGN
population without the limitations in $uv$ plane coverage imposed by
the design of the VLA.

Historically, RLAGN would be selected from a survey and then followed
up with pointed radio observations for detailed study. Here, as with
study at other wavebands, the very large range of angular scales
spanned by RLAGN can be problematic. The closest radio galaxy,
Centaurus A, has scales of interest ranging from tens of degrees to
microarcseconds, and to study even a fraction of that range requires
the combination of data from multiple telescopes \citep{Feain+11}. For
objects at more typical distances the largest angular scales might be
arcminutes, corresponding to hundreds of kpc to Mpc. A correspondingly
large range of interferometer baselines is needed to study the whole
radio structure.

Radio observations at GHz frequencies with sub-arcsecond to arcminute
resolution, as provided by instruments such as the (Jansky) VLA, ATCA and
e-MERLIN, typically correspond to scales of kpc to hundreds of kpc and
are used to study the large-scale structures such as kpc-scale jets,
lobes and hotspots, either in individual objects
\citep[e.g.,][]{Carilli+91,Laing+Bridle02} or large samples
\citep[e.g.,][]{Black+92,Fernini+93,Bridle+94,Fernini+97,Leahy+97,
  Hardcastle+97,Gilbert+04,Mullin+06}. The review of
\cite{Bridle+Perley84} still provides a good summary of many of the
early observational discoveries. Resolved broad-band spectral
mapping gives (model-dependent) information about the age of the
radio-emitting plasma, so-called `spectral ageing'
\citep{Burch77,Myers+Spangler85,Alexander+Leahy87,Harwood+13} based on the
different radiation timescales for electrons of different energies
(see Section \ref{sec:ages}, below). Polarization
imaging at GHz frequencies tell us about the configuration of the
magnetic fields in the large-scale lobes, jets and hotspots
\citep{Laing80,Bridle+Perley84,Laing89,Hardcastle+98} but, because
Faraday rotation effects become dominant at low frequencies,
broad-band polarization studies can also tell us about the thermal
material immediately around or even inside the lobes, both for
individual sources or on a statistical basis
\citep{Dreher+87,Laing88,Garrington+88,Taylor+Perley93,Laing+08,Hardcastle+12,Anderson+18}.

VLBI observations with milliarcsec resolution, corresponding to
physical scales of pc, allow the study of the regions where the jets
are formed and accelerated. Time-resolved studies of jet dynamics are
generally only possible on these scales, and both studies of
individual objects and systematic total intensity studies of large samples
\citep{Lister+16} provide our best constraints on bulk speeds of jets
on these scales. Multi-frequency studies exploiting self-absorption at
lower frequencies can probe magnetic field strengths
\citep{OSullivan+Gabuzda09} while polarization studies constrain the magnetic
field configuration in the jets \citep{Gabuzda+04}.

A good deal of work has been done on the 21-cm line of neutral
hydrogen, either in emission or in absorption against the synchrotron
continuum, in the host galaxies of RLAGN. A key result is that a
number of RLAGN show outflowing neutral hydrogen, presumably
associated with the interaction between the jets/lobes and their
environment, as well as having small-scale nuclear HI components
plausibly associated with the fuel supply. See
\cite{Morganti+Oosterloo18} for a recent review.

\subsection{Optical/IR}
\label{sec:opticalid}

Matching a radio source with an optical counterpart, known as optical
identification, is crucial to any kind of physical interpretation. The
identification of powerful radio sources such as Cygnus A, M87 and
Centaurus A with peculiar galaxies by
\cite{Baade+Minkowski54a,Baade+Minkowski54} marked the beginning of
the study of the physics of these objects, leading, as already noted,
directly to an understanding of the large energies involved, and
thence, through the discovery of quasars \citep{Schmidt63} to the idea
that RLAGN must be powered by accretion onto supermassive
galactic-centre black holes \citep{Lynden-Bell69}.

Optical identification requires the combination of a good radio image
(with resolution sufficient to distinguish between different possible
optical counterparts) and a good optical image, and so it has always
been a significant limitation on the exploitation of radio surveys ---
the final optical identification of the 3CRR sample, for example, came
in 1996 \citep{Rawlings+96}, decades after the initial radio
observations. Deep optical images are required to find the
counterparts of high-redshift radio sources, and these are not easily
available over wide sky areas. This continues to be an issue for
current and next-generation sky surveys, mitigated to some extent by the fact
that the more recent surveys are carried out at an angular resolution
that allows optical identification without requiring followup radio
observations.

Optical counterparts (`host galaxies') of RLAGN have a number of
interesting properties. An important minority are quasars: in other
words, we have a direct view of radiatively efficient nuclear
accretion. Almost all the rest are early-type galaxies, implying a
relationship between RLAGN activity and the most massive systems; but
of those galaxies some, particularly those associated with the most
powerful radio galaxies, are peculiar, showing strong narrow emission
lines similar to those of Seyfert 2 galaxies in optical spectra. The
interpretation of these peculiarities as due to galaxy collision dates
back to the earliest optical identifications
\citep{Baade+Minkowski54a}, and indeed for these powerful objects there
is a higher than average fraction of disturbed or merger-like hosts
\citep{Heckman+86,RamosAlmeida+12}. However, other radio galaxy hosts show no
evidence of peculiarities that cannot be attributed to the jet. We
will return to the implications of these observations for accretion in
Section \ref{sec:fuelling}, while the relationship between radio properties, stellar mass and star formation will be discussed in more detail in Section~\ref{sec:hostenv}.

Optical observations are important in a number of other areas. Optical
synchrotron emission can be identified readily by its polarization,
and it was in fact optical polarimetry that provided early
confirmation of the synchrotron nature of the continuum radiation from
RLAGN \citep{Baade56}, though radio polarimetry soon followed. The
realization that optically-emitting electrons had short lifetimes,
necessitating continuous energy supply, was influential in the
development of jet models for RLAGN. Optical observations of extended
emission-line nebulae around the radio lobes provided early evidence
for the impact of the RLAGN on their environments, e.g.,
\cite{McCarthy+87,O'Dea+02}. Optical studies also provide constraints on the environments of RLAGN --- see Section
\ref{sec:hostenv}.

\subsection{mm/sub-mm/FIR}

RLAGN synchrotron emission persists through to frequencies of 100 GHz
and above and shows little difference from what is seen at lower radio
frequencies \citep[e.g.,][]{Hardcastle+Looney08}. The sub-mm region is important for studies of molecular gas in radio-galaxy environments, fuelling and the interaction of jets with this material. Sub-mm observations demonstrate the presence of molecular gas in the hosts of many nearby radio galaxies \citep[e.g.][]{Prandoni+10,Hamer+14,Rose+19,North+19,Ruffa+19}. In some cases the molecular gas appears mainly located in a rotating disk, while in other cases it may be infalling. As with the neutral hydrogen discussed above, there is also strong evidence in some cases that the molecular material is influenced by the radio lobes
\citep[e.g.,][]{Russell+17,Tremblay+18}. These observations are discussed further in Section~\ref{sec:feedback}. Far-IR observations, probing
the component of dust heated by star formation rather than the AGN,
have shown that nearby normal AGN tend to have low star formation
rates, but the details depend on the type of galaxy
\citep[e.g.,][]{Hardcastle+13} while individual RLAGN associated with very high
dust luminosities and hence, presumably, star-formation rates have
been discovered \citep[e.g.,][]{Barthel+12,Seymour+12}.

\subsection{X-ray}
\label{sec:x-ray}
X-ray observations provide information on the accretion state and
obscuration of the active nucleus, the large-scale components of the
radio source (jets, lobes and hotspots) and the
galaxy-to-cluster-scale hot-gas environment.

Radiatively efficient AGN are strong X-ray sources, and so in many
quasars the X-ray emission is dominated by the active nucleus;
however, at soft X-ray energies this emission is strongly suppressed
by even moderate levels of obscuration, which has the useful effect
that other components of RLAGN can be studied well at these energies.
(We discuss nuclear X-ray emission further in Section
\ref{sec:accmodes}.)

X-ray radiation from the jets, hotspots and lobes is non-thermal and
gives us information about the particle populations responsible for
the emission. A key emission mechanism in this band is inverse-Compton
scattering, where the relativistic electrons responsible for the
synchrotron emission also scatter a photon field to high energies.
Possible photon fields include the CMB, which is always present, but
also synchrotron photons (`synchrotron self-Compton'), starlight from
the host galaxy, radiation from the central AGN itself, or the
extragalactic background light. Where the X-ray emission mechanism is
inverse-Compton scattering, as in the case of lobes \citep{Croston+05}
and some hotspots \citep{Hardcastle+04}, its luminosity depends on the
photon field and on the number of relatively low-energy relativistic
electrons ($\gamma \sim 1000$ for scattering of $z=0$ CMB photons into
the soft X-ray) and so it gives a good constraint on the electron
energetics and therefore, indirectly, on magnetic field strengths and
source dynamics (see Section \ref{sec:cond}). Where instead the
non-thermal X-ray emission mechanism is synchrotron emission, as in
other hotspots and the jets of low-luminosity sources
\citep{Hardcastle+01}, it points to a very energetic population of
electrons (dependent on assumed magnetic field strength, $\gamma \sim
10^8$ or more) which generally implies local ({\it in situ}) particle
acceleration. The prominent jets seen in some high-luminosity objects,
such as 3C\,273 \citep{Harris+Stern87}, PKS 0637$-$752
\citep{Schwartz+00} or Pictor A \citep{Wilson+01} are still the
subject of debate; the original proposal that they represent
inverse-Compton emission from the highly boosted CMB from jets with
$\Gamma \sim 10$ \citep{Tavecchio+00,Celotti+01}, while attractive,
has a number of problems
\citep{Stawarz+04,Hardcastle06,Cara+13,Meyer+15} and it seems likely
that at least some of these powerful jets have a synchrotron origin,
with electron spectra extending through the optical emission region to
the X-ray \citep{Hardcastle+16b}. Jet X-ray properties are reviewed in
much more detail by \cite{Worrall09}.

The thermal X-ray radiation from the hot phase of the ambient medium
of the RLAGN has been of huge importance to constraining dynamical
models of these sources since its first discovery
\citep{Longair+Willmore74,Hardcastle+Worrall00}. More recently,
sensitive X-ray imaging has revealed deficits of X-ray emission
(`cavities') associated with the kpc-scale lobes of many RLAGN
\citep[e.g.,][]{Birzan+04}, and also found a small number of
unambiguous shock features, demonstrating supersonic bulk motion of
the lobes through the medium \citep[e.g.,][]{Kraft+03,Croston+09,Croston+11}. The
implications of these observations are discussed in Sections
\ref{sec:cond} and \ref{sec:dyn}.

\subsection{$\gamma$-ray}
\label{sec:gamma}

Most $\gamma$-ray-emitting RLAGN are blazars, discussed elsewhere in
this volume; in these systems the nuclear $\gamma$ rays are thought to
be Doppler-boosted inverse-Compton scattering of either the
synchrotron continuum itself or of other photon fields (radiation from
the AGN or torus where present --- see Section \ref{sec:accmodes}).
The numbers of non-blazar or so-called misaligned RLAGN that are known
GeV $\gamma$-ray emitters has greatly increased due to the sky survey
of the {\it Fermi} satellite, and now stands at some tens of objects
\citep{Ackermann+15} including well-known nearby radio galaxies such
as Cen A, M87, NGC 1275 and NGC 6251. Of particular interest are the
detections of extended lobes in Cen A and Fornax A
\citep{CenA:Abdo+10,Ackermann+16}. In a leptonic model for the $\gamma$-ray
emission, this can constrain magnetic field strengths just as X-ray
inverse-Compton emission can, but also demonstrates the presence of
high-energy electrons in the lobes. Only a small number of non-blazar
RLAGN (6 at the time of writing: \citealt{Rieger+Levinson18}) are
detected at TeV energies. These again include the nearest powerful
radio galaxies, Cen A and M87. M87's TeV emission is highly variable at
high energies on timescales of days, and appears to be associated with
the sub-pc scale jet \citep{Hada+15}. By contrast, Cen A's emission
seems steady over long time periods \citep{Hess+18} and so an origin
of part or all of it in the large-scale components of the source
remain possible. Inverse-Compton scattering of starlight by the known
population of TeV electrons that produce the X-ray jet is a required
process; if this dominates at high energies then the forthcoming
Cerenkov Telescope Array should able to resolve the jet in Cen A at
TeV energies and permit the first direct magnetic field measurement in
a low-power jet \citep{Hardcastle+Croston11}.

\section{Radio-galaxy populations}

\label{sec:pops}

\begin{figure*}
  \includegraphics[width=\linewidth]{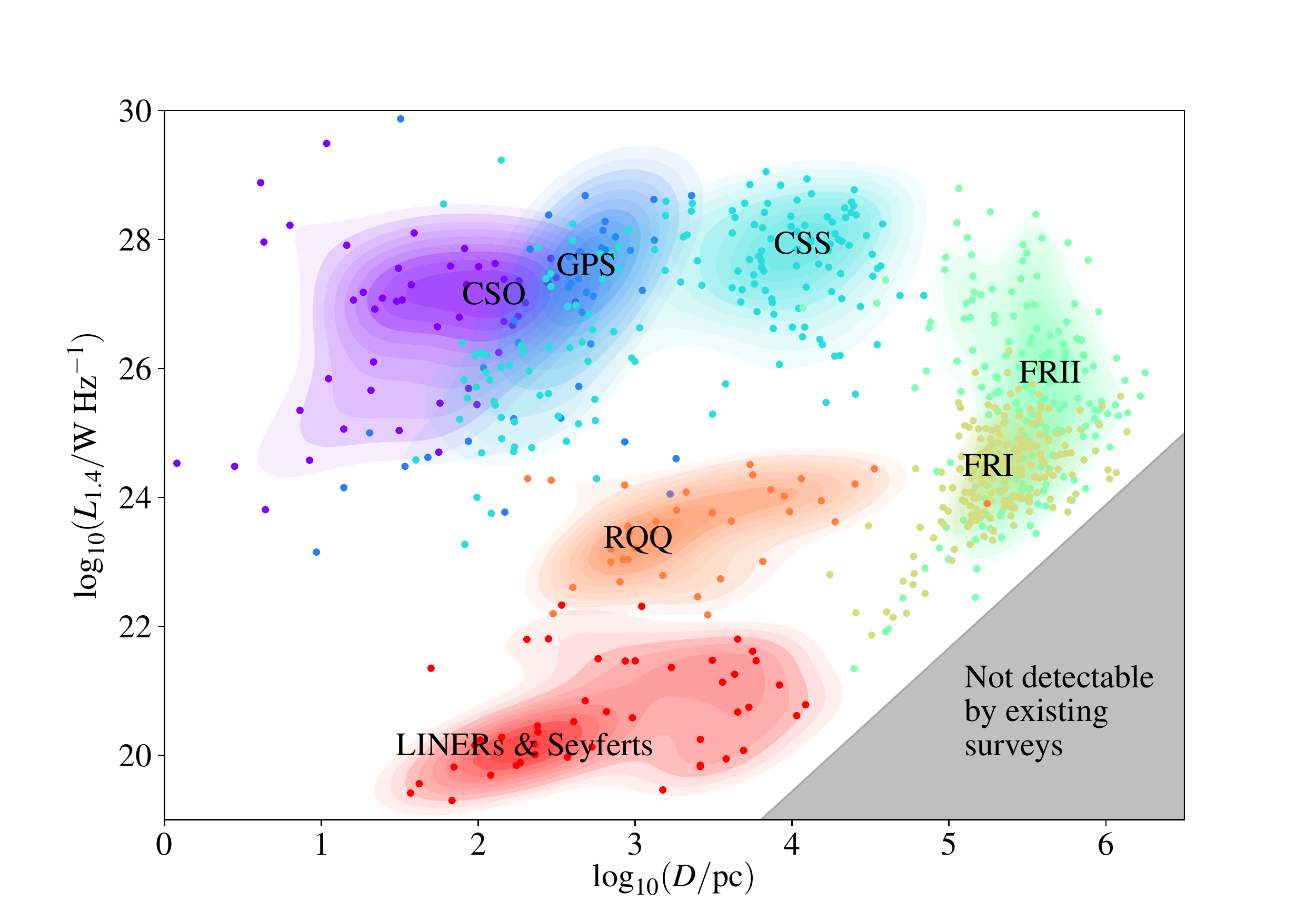}
  \caption{Power/linear-size plot \citep{Baldwin82} for different types of radio-loud
    and radio-quiet AGN, adapted from plots presented by
    \cite{An+Baan12} and \cite{Jarvis+19}. Points show individual
    objects and coloured contours represent a smoothed estimator of
    source density. The different categories of
    source shown are: CSO, GPS, CSS, all objects so classed by
    \citeauthor{An+Baan12}; FRI, FRII, all objects so classed by either
    \citeauthor{An+Baan12} or \cite{Mingo+19} (points are
    representative); RQQ, objects from \cite{Jarvis+19} and
    \cite{Kukula+98}; Seyferts and LINERS, objects from \cite{Gallimore+06};
    and \cite{Baldi+18}. The shaded bottom-right
    corner shows the effect of surface-brightness limitations.}
\label{fig:jarvis_plot}
\end{figure*}

From the first double radio sources in the 3C catalogue to current
low-frequency radio-galaxy samples, our understanding of the diversity
of radio-galaxy populations has evolved dramatically over the last
seventy years.  In Fig.~\ref{fig:jarvis_plot} we summarize the radio
luminosity and size ranges spanned by particular sub-populations of
RLAGN, as discussed in this Section. We note that the samples plotted
in this `$P$-$D$' diagram are all subject to a range of selection
effects, so that gaps between sub-categories do not indicate sharp delineations between separate populations. The main conclusion to be drawn from Fig.~\ref{fig:jarvis_plot} is that AGN-driven jet structures occur over a very wide span in radio luminosity (nearly ten orders of magnitude) and source size (six orders of magnitude). 

In this section we discuss the main classes of radio galaxy that can be categorised primarily from radio observations. We make some reference to host-galaxy and multi-wavelength properties in this section, but we defer full discussion of these topics, and of classification based on nuclear properties, to later sections. We also defer discussion of the cosmic evolution of RLAGN populations to Section~\ref{sec:hostenv}.

\subsection{The Fanaroff-Riley dichotomy}
\label{sec:fr}

The Fanaroff-Riley \citep{Fanaroff+Riley74} morphological distinction
between centre-brightened and edge-brightened radio galaxies (e.g.
Fig.~\ref{fig:agn-picture}), found to be linked to radio luminosity,
has since been widely adopted and applied to many radio catalogues in
the past four decades. While there remains debate about the link
between accretion mode and jet morphology
\citep[e.g.,][and see Section \ref{sec:accmodes}]{Best+Heckman12,Gendre+13,Mingo+14,Ineson+15,Tadhunter16,Hardcastle+07,Hardcastle+09,Hardcastle18b},
the FR morphological divide is thought to be fundamentally linked to
jet dynamics: the edge-brightened FRII radio galaxies are thought to
have jets that remain relativistic throughout, terminating in a
hotspot (internal shock), while the centre-brightened FRIs are known
to have initially relativistic jets that decelerate on kpc scales
\citep[e.g.,][]{Bicknell95,Laing+Bridle02,Tchekhovskoy+Bromberg16}. This
structural difference must necessarily result not purely from
properties of the central engine, but rather from the interplay of jet power and environmental density, so that jets of the same power might in a poor (host-scale) environment remain relativistic and well-collimated, but in a richer environment decelerate, entrain ISM gas, and expand to form turbulent FRI plumes. Such an explanation seemed to find support in the discovery by \cite{Ledlow+Owen96} that the FRI/II luminosity break is dependent on host-galaxy magnitude, so that FRIs are found to have higher radio luminosities in brighter host galaxies (where the density of the interstellar medium is assumed to be higher). However, this result was based on strongly flux-limited samples, with different redshift distributions and environments for the FRIs and FRIIs, and so serious selection effects mean that there is now some uncertainty as to whether this relation in fact holds across the full population of radio galaxies \citep{Best09, Lin+10, Wing+Blanton11, Singal+Rajpurohit14, Capetti+17, Shabala18}.

As radio surveys have reached lower flux limits,
evidence has also emerged that the FR morphological division is less closely tied to radio luminosity than was previously thought, with the emergence of an unexpected population of low-luminosity sources with
edge-brightened FRII morphology \citep{Best09,Miraghaei+Best17,Capetti+17,Mingo+19}. Using a
large sample from the LOFAR Two-Metre Sky Survey (LoTSS),
\cite{Mingo+19} have shown that these `FRII-low' radio galaxies
form a substantial fraction of the FRIIs at $z<0.8$ (see also Fig.~\ref{fig:jarvis_plot}). This population was absent from early studies as they are rare in the very local Universe, and hence have fluxes below the limits of early surveys such as 3CRR \citep{Mingo+19}. The substantial overlap in luminosity range for FRIs and FRIIs found
in modern samples does not necessarily break the jet disruption
paradigm for the morphological distinction. Firstly, the conversion
from jet power to radio luminosity has very large scatter and
systematic biases, as discussed in Section \ref{sec:power}
\citep{Hardcastle18,Croston+18}. Secondly, there does appear to be a
link between morphology and host brightness \citep{Mingo+19} so that
the low-luminosity FRIIs may occupy particularly poor local
environments, consistent with their remaining undisrupted despite
their (presumably) comparatively low jet power. 

\subsection{Blazars}

Blazars are a sub-category of radio-loud AGN with distinct properties,
including bright variable emission at a range of wavelengths, thought
to reflect relativistic effects in a jet oriented at a small angle to
the line of sight. Blazars show a diverse range of radio structures
\citep[e.g.,][]{Rector+Stocke01,giroletti+04}. It is thought that
sub-classes of blazars can be unified with radio galaxies and quasars,
with apparent difference in properties explained by orientation
effects (Section \ref{sec:accmodes}). We do not discuss blazars or the
consequences of their properties for source models in detail in this
work, deferring to the dedicated Chapter in this volume.

\subsection{Cluster radio-galaxy populations}
\label{sec:cluster}

The large-scale environments of radio galaxies are discussed in Section~\ref{sec:hostenv}. The majority of radio galaxies do not live in rich cluster environments; however, those radio galaxies that do inhabit the richest environments show a range of characteristic morphological features, likely to be caused by their dense surrounding medium \citep[e.g.,][]{Owen+Ledlow97}. 

While there are examples of cluster-centre radio galaxies with
morphologies typical of the FRI or FRII class (e.g., Cygnus A), many
cluster-centre radio galaxies, including the most famous
cluster-centre radio galaxy Perseus A (3C\,84), show amorphous radio
structure with no evidence for collimated jets on scales of tens of
kpc \citep[e.g.,][]{Miley+Perola75,Burns90,Owen+Ledlow97}. Another
radio-galaxy sub-population that is strongly associated with galaxy
clusters are the bent-tail sources, which can be further separated
into the head-tail / narrow-angle (NAT) and wide-angle tail (WAT)
sources \citep{Owen+Rudnick76,O'Dea+Owen85}. The bent jets and
plumes/tails of these sources are thought to be left behind as the
host galaxy moves with respect to the intracluster medium, and have been the subject of extensive dynamical studies. In Section~\ref{sec:hostenv} we comment on their relation with their --- typically cluster --- environments, and in Section~\ref{sec:uses} discuss their use as probes of dense environments in the more distant Universe.

As lower-frequency radio observations have become increasingly
sensitive, it has become clear that another common feature of cluster
centre radio sources is the presence of low surface brightness
extended lobes permeating a larger volume than the currently active
source, and likely to indicate previous episodes of activity --- in
some cases linked to the presence of outer `cavities' (surface
brightness depressions) in the X-ray emission from the intracluster
medium. The nearby radio galaxy M87 is an example of a cluster source
with a very extended low-surface brightness halo pervading the
intracluster medium \citep{Owen+00,DeGasperin+12}. Galaxy clusters
also frequently possess diffuse radio emission that is not directly
associated with current radio jet activity (haloes and relics)
\citep{Feretti+Giovannini96}. The links between past and current radio
galaxy activity and cluster diffuse radio emission are an interesting
topic of current research, which is discussed further in
Section~\ref{sec:uses}.
 
\subsection{Restarting and remnant radio galaxies}
\label{sec:remnant}

Double-double radio galaxies \citep{Schoenmakers+00} and other restarting radio galaxies \citep[e.g.,][]{Clarke+Burns91,Bridle+89,Jamrozy+07} are further populations that provide important insights into radio-galaxy life cycles and triggering. Spectral and morphological studies have also been used to identify `dying' radio galaxies \citep[e.g.,][]{Murgia+99}, now commonly known as remnant radio galaxies (we avoid the term `relic' radio galaxy in this context, to avoid confusion with cluster radio relics). 
Double-double radio galaxies (DDRGs) are systems in which an inner pair of radio lobes propagate along the same axis as outer lobes. Typically, both sets of structure have edge-brightened, FRII-like morphology. Samples of tens of DDRGs now exist, from investigations with the FIRST, NVSS, and most recently LoTSS surveys \citep{Nandi+Saikia12,Kuzmicz+17,Mahatma+19}. \cite{Mahatma+19} investigated the host galaxies of a sample of DDRGs, finding no significant differences to a control sample of similar luminosity ordinary radio galaxies. It appears that DDRG structure is not caused by specific host galaxy conditions, but is likely to relate to accretion conditions being interrupted and restarting. There have also been detailed studies of other types of candidate restarting objects, in which small-scale sources are surrounded by extended halos and/or misaligned outer lobes \citep[e.g.,][]{Jamrozy+07,Jetha+08,Brienza+18}. The recent LoTSS study of \cite{Mingo+19} suggests that there may be a moderate sized population of newly restarted jets embedded in extended radio emission, but follow-up work is needed to investigate this population further. Larger statistical samples and spectral modelling are needed to draw firm conclusions about the prevalence and duty cycle of recurrent activity, but it is clear that objects with visible signatures of recurrent activity are rare in the radio-loud AGN population.

Remnant radio galaxies can be difficult to identify due to their low
surface brightness, and until recently relatively few were known
\citep[e.g.,][]{Parma+07,Murgia+11,Saripalli+12}. Sensitive
low-frequency surveys were expected to turn up larger numbers of these
sources, expected to be dominated by steep-spectrum aged plasma.
\cite{Brienza+17} identified 23 candidate remnant sources in the
Lockman Hole survey area, while \cite{Mahatma+18} searched for radio
cores in a low-frequency selected sample, finding that only 11/33
candidate remnant sources showed no radio core, indicating a true
remnant (note that some sources with a radio core may be restarting radio galaxies). Population modelling and theoretical work indicate that the observed remnant fraction should be low, as lobes will fade through radiative losses on a short timescale after the jet turns off \citep{Godfrey+17,Hardcastle18b}. 

\subsection{Compact and unresolved radio-loud AGN}

In addition to the extended radio galaxy populations, populations of
compact radio galaxies, including luminous populations of compact
symmetric objects (CSOs), gigahertz-peaked spectrum (GPS) sources, and
compact steep spectrum (CSS) sources have been studied for many decades \citep[e.g.,][]{O'Dea+Baum97,O'Dea98}. More recently fainter populations of compact sources have been shown to constitute a very large AGN population \citep[e.g.,][]{Sadler+14,Baldi+15,Gurkan+18,Hardcastle+19}, thought to possess low-power small-scale jet activity. These compact radio-loud AGN are of importance for understanding radio-galaxy triggering and life cycles. We briefly discuss the main subclasses before commenting on current understanding of their relationship(s) to the large-scale radio galaxy population.

Compact steep-spectrum and gigahertz peaked spectrum sources are
powerful radio-loud AGN ($L_{\rm 1.4GHz} > 10^{25}$ W Hz$^{-1}$)
smaller than 1 -- 2 arcsec \citep[e.g.,][]{O'Dea98,Orienti16}. The CSS
sources have typical physical sizes extending to $\sim 20$ kpc, while
the GPS sources are smaller, typically $<1$ kpc. Compact symmetric objects (CSOs) are the smallest sources, with sizes of less than a few hundred pc. The radio spectra of
CSS sources peak at MHz frequencies, while the higher-frequency GHz turnover in GPS
sources is traditionally thought to indicate self-absorption due to high densities
in very compact synchrotron-emitting regions, although free-free
emission is another possible mechanism to explain the observations. CSS sources possess
interesting features in optical emission line and molecular gas,
indicative of jet/environment interaction on small scales. Recently,
\cite{Callingham+17} have presented the largest sample to data of
such systems, obtained with the MWA GLEAM survey. There have been two
competing hypotheses for these populations: that they are `young'
sources that will evolve to become traditional FRI or FRII radio
galaxies \citep{O'Dea98}, or that they are `frustrated' jets
occurring in dense environments and unable to grow to a large size
\citep[e.g.,][]{vanBreugel+84}. It is likely that both scenarios are
  relevant for a subset of objects, since small, young objects will
  necessarily probe the densest possible parts of a particular
  environment. In principle, low-frequency spectral information may be
  able to distinguish the influence of synchrotron self-absorption and
  free-free absorption, expected in a dense environment ---
  \cite{Callingham+17} found a small sample of potential candidates
  in which free-free absorption may be important. Observations of X-ray absorption can also be used
  to search for a dense medium around compact sources
  \citep{Sobolewska+19}.

As shown in Fig.~\ref{fig:jarvis_plot}, the CSO, GPS and CSS categories have historically applied to objects of high luminosity, due to the high flux limits of early radio surveys. However, in addition to the peaked spectrum and luminous compact AGN
populations, there is a large population of low-luminosity radio
sources with evidence for small-scale jets. Radio nuclei in ordinary
elliptical galaxies have been known for many years \citep[e.g.,][]{ho99},
and have recently been investigated systematically at high resolution
\citep{Baldi+08}. Low-luminosity kpc-scale jet structures have been studied
extensively in Seyfert galaxies
\citep[e.g.,][]{Gallimore+06,Hota+Saikia06,Croston+08a,Mingo+11,Jones+11,Mingo+12,Williams+17},
and examples of small-scale jets are emerging in radio-quiet quasars
\citep[e.g.,][]{Jarvis+19}.

Recently the `FR0' nomenclature has been introduced to describe the
population of unresolved low-luminosity radio-loud AGN
\citep[e.g.,][]{Sadler+14,Baldi+15,Hardcastle+19}. We do not favour
this terminology: as stated by \cite{Fanaroff+Riley74}, the FRI/FRII
classification is a morphological one, and as such it relies on
observations capable of resolving the source (i.e.\ the source must be
at least a factor of a few larger than the beam size of the radio
observations before a classification can be made). If these
observations do not exist, the FR class of the source is unknown ---
this gives rise to the large fraction of sources classed as `C'
(`compact') in early catalogues such as the 3CRR catalogue of
\cite{Laing+83}. When higher-resolution observations become available,
as is now the case for the 3CRR objects, the sources can be
classified. A classification based on a source's unresolved nature in
a particular survey (with a particular resolution and surface
brightness limit) can never be physical, and indeed some evidence
exists that objects selected this way are a heterogeneous population
\citep{Baldi+19}. Nevertheless, it is certain that this dominant population
of compact radio-loud AGN in the nearby Universe --- however they are
designated --- are of great importance for our understanding of
radio-galaxy life cycles and AGN feedback, and are in need of further
study.

\section{Physical conditions}

\label{sec:cond}

A great deal of observational work has gone into measurement of the
physical conditions in various components of the radio sources and in
the external medium. Below we summarize some of the current
understanding on this topic.

\subsection{Jet speeds}
\label{sec:jetspeed}

VLBI observations of pc-scale jets measure apparent superluminal
motions, $v_{\rm app} > c$. These imply highly relativistic jet bulk
motions on pc scales: superluminal apparent speeds as high as $50c$
have been observed \citep{Lister+16}, which on standard assumptions
(in particular equating the pattern speed, or the speed of motion of
structures in the jet, with the bulk flow speed) would imply bulk
Lorentz factors $\Gamma >50$. These very high values may represent the
extremes of a dispersion of bulk speeds within a given source, in
which case we would expect internal dissipation to reduce the
effective bulk speed on scales of hundreds of pc. However, it seems
clear that the jet may leave the inner parts of the radio source with
highly relativistic bulk speeds.

These speeds are in contrast to those estimated using Doppler boosting
arguments from unified models of RLAGN (see Section
\ref{sec:accmodes}). Constraints derived from luminosity functions of
blazars and their presumed parent population of radio galaxies
\citep{Urry+91,Hardcastle+02} imply bulk beaming speeds of $\Gamma \sim
3$--$5$. More problematically, all estimates of the kpc-scale jet
speeds in powerful FRII RLAGN tend to give $\beta \approx 0.5$ --
$0.7$
\citep{Wardle+Aaron97,Hardcastle+99,Arshakian+Longair04,Mullin+Hardcastle09};
these estimates rely on the demonstration using polarization
statistics in quasars \citep{Laing88,Garrington+88} that the
one-sidedness of kpc-scale jets is indeed due to beaming. With the
debatable exception of the beamed inverse-Compton model for powerful
jets (Section \ref{sec:x-ray}) there is no direct evidence for high
bulk speeds on hundred-kpc scales in powerful objects, but there is
also no evidence for the strong deceleration that would be required to
account for the differences between pc and kpc scales. A plausible
explanation is that the jet is structured and that most of the
emission seen on kpc scales comes from a slow-moving component while
most of the energy is carried by a relativistic outflow. There is some
direct evidence for this picture in the shape of edge-brightened
structures of a few resolved powerful jets
\citep{Swain+98,Hardcastle+16b} but more observational work is needed.

In FRI radio galaxies the situation is clearer. These objects often
show jets that are one-sided on small scales but two-sided on large
scales, implying bulk deceleration if the sidedness is attributed to
Doppler boosting. Because the jets are resolved both transversely and
longitudinally, detailed models of the jet velocity field of
individual objects can be constructed, using the polarization of the
jets to break degeneracies involving the unknown angle of the jet axis
to the line of sight. The results
\citep[e.g.,][]{Laing+Bridle02,Laing+Bridle14} show that jets
initially have $\beta \sim 0.8$ on the kpc scale and decelerate
smoothly (but often more rapidly at the edges) to a constant,
sub-relativistic speed. The bulk deceleration must involve entrainment
of material initially at rest with respect to the jet
\citep{Bicknell94}, but it is not clear what this material is: stellar
winds may provide some of the required mass \citep{Bowman+96,Wykes+15}
but if this is insufficient, as it probably is in the most powerful
sources, then entrainment of the external medium in some form may be
required. It is important to note that the {\it lobe dynamics} of FRIs
should not in themselves be affected by this entrainment --- the
momentum and energy flux of the jet are essentially unchanged by
it.

\subsection{Environments}

For the purposes of dynamical modelling, profiles of the pressure and
density in the external medium are required. These can be derived from
deprojection (normally spherical deprojection) of the observed X-ray
surface brightness profile if we assume that the X-ray-emitting gas
dominates density and pressure ---  an assumption that is more valid on
larger scales than smaller ones --- and from X-ray spectroscopy to
estimate the temperature (profile) of the hot gas. Bulk inference of radio galaxy
environments from X-ray data is now relatively routine (see e.g.,
\cite{Worrall+Birkinshaw00,Hardcastle+Worrall00,Croston+08,Ineson+15}) though systematic
uncertainties may arise through the assumption of spherical symmetry.
It is far more difficult to estimate the environmental density and
pressure profiles based on optical information alone.

\subsection{Lobe energy density and particle content}
\label{sec:bfield}

Traditionally estimates of the lobe energy density have involved the
minimum-energy condition \citep{Burbidge56} (or, equivalently,
equipartition of energy between electrons and magnetic field).
Interpreted as lower limits, these are valid, but they do not give
many constraints on source dynamics. In minimum-energy calculations in the literature widely differing
assumptions are often made about the energetic contribution of
non-radiating protons, and care is also needed to distinguish between
calculations made using a minimum and maximum {\it frequency} for the
radiating particles (which implies field-dependent energy integration
limits) and one using a minimum and maximum {\it energy} or Lorentz
factor. \cite{Beck+Krause05} discuss some of these issues in detail.

A better approach, where possible, is to infer magnetic field strength
and so energetics from observation. Observations of inverse-Compton
emission from the lobes and hotspots of FRII RLAGN
\citep{Hardcastle+02,Hardcastle+04,Kataoka+Stawarz05,Croston+05,Ineson+17}
measure the number of inverse-Compton-scattering electrons and so
allow the characteristic magnetic field strength to be estimated based
on the observed synchrotron emission. This gives typical field
strengths of a factor $\sim 2$--$3$ below the equipartition values and
so energy densities in field and electrons a factor 2.5--5 higher than
the equipartition value. On these assumptions radio lobes tend to be
strongly overpressured at the radio lobe tip (with respect to the
environment at that location) and moderately overpressured at the
mid-point of the lobe. It is then possible to argue
\citep{Hardcastle+02,Croston+18} that protons cannot be strongly
energetically dominant over electrons in these lobes, as, if they
were, this good agreement over a number of objects between the inferred internal pressures and
observed external pressures would be a coincidence. This argument does not rule out a contribution to the energy
density from protons that is comparable to that of the electrons or
the magnetic field; there is some evidence that FRIIs in rich
environments may have a higher non-radiating particle content
\citep[e.g.,][]{Hardcastle+Croston10}. It has long been known from
low-frequency polarization observations that the number density of internal
{\it thermal} electrons in FRII lobes must be relatively low
\citep{Scheuer74}, consistent with the idea that their internal
pressure is dominated by the radiating particle population.

The situation is more complicated for the FRI radio galaxies for two
reasons. Firstly, there are few direct measurements of X-ray
inverse-Compton from the lobes that would allow a magnetic field
measurement, as the thermal emission tends to dominate over
inverse-Compton \citep{Hardcastle+Croston10}; secondly, a comparison
between radio emission and external pressure often requires departures
from equipartition much more substantial than those measured in the
FRII population
\citep{Morganti+88,Killeen+88,Feretti+90,Taylor+90,Feretti+92,Bohringer+93,Worrall+95,Hardcastle+98,Worrall+Birkinshaw00,Hardcastle+05,Croston+03,Dunn+05,Croston+18}.
In some cases these highly sub-equipartition fields are ruled out by
the lack of observed inverse-Compton emission in the X-ray.
Observations of cavities (Section \ref{sec:cluster}) and of
low-frequency polarization from the lobes rule out a
model in which the lobe is filled with thermal particles similar to
those in the external medium. The most plausible hypothesis is that an
additional population of high-energy particles dominates the lobe
energetics in the FRIs, but not the FRIIs. Relating this to the other
known difference between the two populations, we can hypothesise that
the material that is entrained as the FRI jets decelerate is then
heated to provide the missing pressure \citep[e.g.,][]{Croston+Hardcastle14,Croston+18}. This has
the effect that FRI radio galaxies, where the lobe dynamics are
dominated by an invisible, non-radiating particle population, are much
more difficult to model than the higher-power but less numerous FRIIs.
A prediction of this model is that internal depolarization, due to
entrained thermal electrons, might be visible in lobes of low-power
sources, and broad-band polarimetry is starting to probe this regime \citep[e.g.][]{Anderson+18}

Finally, it is worth noting that there are as yet few direct
measurements of lobe kinematics. For the smallest class of double
radio sources known, the CSOs, the lobe expansion speeds in the plane
of the sky can be measured directly using VLBI
\citep[e.g.,][]{Owsianik+Conway98,Polatidis+Conway03,Gugliucci+05},
giving sub-relativistic expansion speeds of order a few tenths of $c$.
A few objects drive observable shocks into the external medium which
can give an instantaneous estimate of the lobe advance speed if the
properties of the unshocked medium are known
\citep{Croston+07,Croston+09,Croston+11,Snios+18}; however, numerical
modelling of the shocked shells suggests that it is not trivial to
interpret these measurements. Next-generation X-ray satellites such as
{\it Athena} \citep{Nandra+13} will provide measurements of shock
properties for large samples of radio galaxies. For larger sources, a
traditional method to estimate lobe advance speeds statistically is to
consider lobe length asymmetries in complete (low-frequency selected)
samples \citep[e.g.,][]{Longair+Riley79,Scheuer95} since light travel
time effects from the nearer to the further lobe will give an apparent
difference in lobe length between the two which depends on the lobe
advance speed. However, this method is biased if it is assumed that
the lobe length difference arises purely from light travel time
effects, since environmental effects may also play a role in lobe
asymmetry (i.e. the longer lobe may not always be the nearer). If the
lobe containing the (brighter) jet, rather than the longer lobe, is taken to be
the nearer lobe in order to reduce this bias, as was done by
\cite{Scheuer95} (relying on the assumption that kpc-scale jets are
beamed as discussed in Section \ref{sec:jetspeed}), the method gives lobe advance speeds of
order a few per cent of the speed of light. These would still be
highly supersonic, ${\cal M} = 10$, with respect to the sound speeds
even in a rich cluster medium. By contrast, the pressure ratio
estimates of \cite{Ineson+17} (which however exclude the ram pressure
of the jet) suggest a median ${\cal M} \sim 2$. It is not as yet clear
what causes the discrepancy.

\subsection{Lobe ages}
\label{sec:ages}

Estimating the ages of RLAGN is difficult. The one exception is where
a direct estimate of the current source expansion speed exists; then a
simple `kinematic age' can be calculated on the (clearly incorrect but
not badly wrong) assumption of constant expansion speed. This has been
done using observed proper motions in CSOs, as described in the
previous Section, yielding ages of the order of $10^3$ years for these
smallest of double-lobed objects. For larger objects a `dynamical age'
can be estimated based on the projected source size and estimated
expansion speeds as described above --- this would lead to estimates
of dynamical age of between 10 and 100 Myr for a source with a lobe
100 kpc in length. However, as noted above, these speed estimates are
very uncertain. A better approach to dynamical age estimation, where
possible, is to fit some source model (see the next Section) to the
observed radio properties of the source and optimize for dynamical
age.

A very popular alternative approach is to estimate ages by `spectral
ageing' \citep[e.g.,][]{Myers+Spangler85}, in which the energy-dependent loss rate of
synchrotron-emitting electrons is used to estimate the age of the
source (and/or, since it gives ages for all positions inside the lobe, the
speed of the internal transport of the radiating plasma). This method
tends to give ages of the order of $10^7$ years for 100-kpc-scale
lobes \citep{Alexander+Leahy87,Harwood+13} and there is therefore a
discrepancy of up to an order of magnitude (in the sense that the
spectral age is lower) between estimates of the spectral and dynamical
ages. Although the spectral age is model-dependent, with the greatest
difference being between models that assume effective pitch angle
scattering of the radiating electrons \citep{Jaffe+Perola73} and those
that do not \citep{Kardashev62,Pacholczyk70}, the derived ages are generally
similar whichever model is used. This problem has been apparent for many years \citep{Eilek96}
and at the time of writing the solution seems likely to be a
combination of several factors. Firstly, we now know (see above,
Section \ref{sec:bfield}) that the
equipartition field strengths used in most spectral ageing studies
overestimate the field strength by a factor of a few, and consequently
these studies underestimate the spectral age, except in situations where
inverse-Compton losses to the CMB dominate over synchrotron losses.
Spectral ageing studies actually measure the break frequency $\nu$, and this
is related to the age $t$ by
\begin{equation}
t = \left({{CB}\over{(B^2+B_p^2)^2\nu}}\right)^{1\over 2}\label{lifetime}
\end{equation}
where $B$ is the magnetic field strength, $B_p$ is the equivalent
magnetic field to the energy density in photons $u_p$ (in SI units, $B_p =
\sqrt{2\mu_0 u_p}$; this term describes losses due to inverse-Compton
scattering of background photons) and $C$ is a constant. For $B \gg B_p$,
$t \propto B^{-3/2}$ and so decreasing $B$ can increase the estimate
of $t$. Secondly, it remains possible that there is some {\it in situ}
particle acceleration in the lobes, which would tend to reduce the
observed spectral ages: indeed such distributed (leptonic) particle acceleration must take place at
some level in models in which RLAGN are responsible for the
ultra-high-energy cosmic ray population (Section \ref{sec:uhecr}).
Thirdly, the mixing of old and young electron populations within the
lobes can in principle allow a source to continue to show an apparent
age gradient while reducing the apparent maximum age, as shown
recently by \cite{Turner+18}. It seems plausible that a combination of
these, particularly the first and third, can bring spectral and
dynamical ages into agreement \citep{Mahatma+20}.

It is finally worth noting that in the cavity power method of
estimating jet powers, which we discuss further in the subsequent
section, lobe ages are estimated from either the sound crossing time,
the buoyant rise time, or the `refilling timescale' of the cavity
\citep{Birzan+04}: of these the sound crossing time of the source is
generally the shortest. But for a typical powerful source, we expect
the lobe expansion to have been supersonic for a significant fraction
of the source's lifetime, and thus these age estimates are likely to
be significant overestimates with respect to a true dynamical age.

\subsection{Jet power}
\label{sec:power}

Given the uncertainties on both lobe energetic content (Section
\ref{sec:bfield}) and age (Section \ref{sec:ages}) it will come as no
surprise that estimates of the lifetime-averaged jet kinetic power,
which we denote $Q$, are also highly uncertain. Basic calculations
from the lobe energetics and lifetimes given above for large sources
suggest that jet powers should lie in the range $10^{36}$ -- $10^{40}$
W, but these do not take into account the work done on the external
medium. Other than the case of a small number of FRIs with
well-modelled jet dynamics \citep[e.g.,][]{Laing+Bridle02} the
situation is best for the powerful FRII-type sources where the lobe
energetics are well constrained and dynamical models are well
understood. It is then possible to try to relate observable quantities
like radio luminosity to the jet power by way of a model
\citep{Willott+99}, making some assumptions about the energy transfer
to the source environment. In practice, this is simplistic, since we
know that the radio luminosity for a source of constant jet power must
vary not just with time but also with the environment of the radio
source \citep{Hardcastle+Krause13} and with redshift, since
inverse-Compton losses are redshift-dependent. More sophisticated
approaches to jet power inference try to take these effects into
account \citep{Hardcastle+19}.

An alternative approach is to construct jet power estimates from X-ray
observations of cavities excavated by the lobes in the external
medium, estimating the $p\Delta V$ work done to inflate the cavity
using estimates of pressure derived from fitting to the X-ray
spectrum. This has the advantage that a quantity directly related to
the work done on the external medium can be calculated. However, it
only works when cavities are observed, which rules its use out in the
case of the most powerful `classical double' AGN, where typically the
lobes are brighter in inverse-Compton than their surroundings
\citep{Hardcastle+Croston10}; these sources also generally drive
shocks into the external medium which are not accounted for in the
$p\Delta V$ calculation. The method is biased towards small
sources in rich cluster environments \citep{Birzan+12} and relies on
expensive X-ray observations that are not available for large samples
of sources. As a method for calculating jet power, it also relies on
poorly known source ages, as discussed above. For all of these
reasons, though the jet power estimates from this methods do seem to
show some correlation with radio luminosity \citep{Cavagnolo+10}, the
scatter and biases mean that any inference of a jet power from the
calibration of such a relation must be done with extreme caution.
Indeed, \cite{Godfrey+Shabala16} suggest that there is no physically
meaningful correlation between radio luminosity and calculated jet power for these objects at all. Whether this
is the case or not, more sophisticated power inference methods are likely to be needed in future.

Observations of restarting radio galaxies (Section \ref{sec:remnant})
imply that it is possible for the jet power to drop to very low levels
(perhaps, but not necessarily, to zero) and then to recover on a
timescale shorter than the overall source lifetime. We have very
little information on the power spectrum of such variations in a
typical source, and estimates based on the large-scale structure or
the impact on the large-scale environment can only really provide a
lifetime-averaged rather than an instantaneous value.

\section{Dynamics: modelling and simulation}
\label{sec:dyn}
The basic dynamical picture of RLAGN has remained unchanged since the
work of Scheuer, Blandford and Rees in the 1970s
\citep{Scheuer74,Blandford+Rees74}. Oppositely directed light jets,
sometimes called beams, are emitted from the central engine and impact on the
external medium. Because the jets are light compared to the external
medium, conservation of momentum dictates that the flow speed up the
jet must be much higher than the advance speed of the contact surface,
so the conditions are right (assuming that cooling is slow) for
material to be redirected away from the head of the source and flow
sideways and backwards to form lobes. Once lobes are present, their
expansion is governed by their internal pressure as well as by the ram
pressure of the jet material. If/when the expansion of the lobes is
supersonic, they will drive a shock into the ambient medium, so we
can then consider three zones of interest; the lobes themselves, the
shell of swept-up, shocked gas around them, and the undisturbed
external medium which has not so far been affected by the radio
source. \cite{Scheuer74} introduced an important variation on this
model: once the lobes are no longer strongly overpressured with
respect to the external environment, they will cease to drive shocks
in the transverse direction, and may eventually be squeezed outwards
away from the central engine by the pressure of the external medium.

Modelling of radio galaxy lobes is simple in principle but complex in
detail, particularly if we wish to use even modestly realistic
environments. Early modelling, such as that of \cite{Scheuer74}, laid
out the basic principles of lobe dynamics in a uniform atmosphere. The
important work of \cite{Kaiser+Alexander97} used a power-law
atmosphere and, considering the collimation of the jet, derived
equations for self-similar growth of radio sources that have been
widely used (they were followed in some of these assumptions by, e.g.,
\citealt{Kaiser+97}, \citealt{Blundell+99}, \citealt{Nath10},
\citealt{Mocz+11} and \citealt{Godfrey+17}) and allowed the prediction
of source evolutionary tracks in a power/linear-size diagram like that
of Fig.\ \ref{fig:jarvis_plot}. But in fact radio galaxy atmospheres are
not scale-free power laws, but have a scale that relates to the mass
of the halo \citep{Arnaud+10,Sun+11}, which invalidates the
assumptions of self-similar lobe evolution. Moreover, the assumptions of the model
of \cite{Kaiser+Alexander97} also restrict it to the case where the
source remains strongly overpressured at all times, which is a
self-consistent requirement \citep{Begelman+Cioffi89} but not
obviously observationally the case. \cite{Hardcastle+Krause13}
investigated a self-collimating jet model comparable to that of
\cite{Kaiser+Alexander97} in numerical simulations and showed that in
reasonably realistic $\beta$-model environments the self-similarity
assumptions do not hold; sources show time-variable axial ratios in
the sense that the ratio of lobe width over lobe length gets lower
with time, and at late times the lobes come into transverse pressure
balance with the external medium are pushed away from the central part
of the host environment by buoyancy forces, as originally proposed by
\cite{Scheuer74}. On the other hand, analytical models such as those
of \cite{Luo+Sadler10} which assume expansion in pressure balance do
not seem likely to be realistic for young sources which must start out
(and observationally are) overpressured on any assumption. For these
reasons more recent analytical modelling tries to capture the insights
from numerical models and deal with the evolution of the sources from
the overpressured to the pressure-balanced phases. Examples of this
more recent approach are \cite{Turner+Shabala15} and
\cite{Hardcastle18}; these models predict, for example, evolution of
the radio luminosity with time (or, equivalently, source size) that is
qualitatively consistent with the results of numerical modelling. None
of these models currently deals well with the possibility of strongly
time-varying $Q$ (Section \ref{sec:power}) or of changes in the jet
direction over time, e.g. due to precession induced by a close binary
black hole pair \citep{Krause+19}.

A vast body of work exists on numerical modelling of the large-scale
structure of radio sources and on their impact on the external medium;
some of this will be discussed elsewhere in this volume. The value of
numerical modelling has been recognised since the very earliest
simulations \citep{Norman+82,Williams+Gull85} but there are many
difficulties in carrying out detailed simulations of RLAGN, including
the very large spatial dynamic range (in principle from the jet
generation scale to the Mpc scales of the largest lobes), the fact
that relativistic bulk motions and non-negligible magnetic fields are
both expected to be present, the fact that radio sources are clearly
not axisymmetric so that three-dimensional modelling is needed, the
difficulty of accurately modelling particle acceleration, transport
and radiative losses and, at least in the early days of modelling, the
very poorly known physical conditions in the lobes and environment
(see previous Section).

Some approaches to simulations, which emphasise different parts of
this parameter space, include:
\begin{enumerate}
  \item Trying to reproduce large-scale lobe dynamics, often in the
    context of parameter studies varying jet and environmental
    properties
    \citep[e.g.,][]{Norman+82,Cioffi+Blondin92,Massaglia+96,Carvalho+O'Dea02,Krause03,ONeill+05,Hardcastle+Krause13}.
    These models often omit relativistic effects and magnetic fields and neglect particle
    acceleration or transport, and may be simplified by the assumption
    of axisymmetry. They reproduce the inflation of lobes
    by light (but not heavy) jets and can generate internal jet
    termination shocks which are assumed to be related to the hotspots
    observed in FRIIs. As noted above, departures from self-similarity
    are observed in realistic environments.
\item Trying to reproduce detailed features of jets or lobes in total
  intensity or polarization
  \citep[e.g.,][]{Bodo+98, Rossi+08, Perucho+10,Mignone+10,Gaibler+09,Huarte-Espinosa+11,Hardcastle+Krause14}. In this
  case relativistic effects are often important and magnetic fields
  may well be modelled, but large volumes and realistic environments
  are less important.
\item Modelling of particle acceleration, transport and loss
  \citep[e.g.,][]{Jones+99,Tregillis+01,Tregillis+04,Mendygral+12,Vaidya+18}.
  These models can qualitatively reproduce many of the complex
  features seen in the synchrotron emission of real sources, and also
  the effects of radiative losses on radio spectra.
\item Modelling of the impact of RLAGN on their environments. The
  impact of powerful sources on the hot gas has been particularly well
  studied
  \citep[e.g.,][]{Basson+Alexander03,Zanni+03,Omma+Binney04,ONeill+05,Gaibler+09,Hardcastle+Krause13,Bourne+Sijacki17}
  but more recently interaction with cold gas, which may be important
  in small-scale sources or at high redshift, has also been modelled
  \citep{Sutherland+Bicknell07,Gaibler+11,Gaibler+12,Mukherjee+18}.
\end{enumerate}
  
\section{Central engines}

\subsection{Unified models and accretion modes}
\label{sec:accmodes}

Observations of the optical spectra of radio galaxy hosts
\citep{Hine+Longair79,Laing+94} show a wide range of possible optical
behaviour. While some objects show strong high-excitation broad and
narrow lines similar to those in Seyfert galaxies, others exhibit weak
or no line emission. This observational dichotomy has been given a
number of names but here we begin by following \cite{Laing+94} in separating RLAGN
observationally into `low-excitation radio galaxies' (LERGs) and
`high-excitation radio galaxies' (HERGs). The latter class includes
both `narrow-line radio galaxies' (NLRGs) and `broad-line radio
galaxies' (BLRGs) which have optical spectra resembling those of
classical Seyfert 2 and Seyfert 1 galaxies respectively, as well as
optically selected quasars, which have spectra similar to those of the
BLRGs but by definition show dominant optical continuum emission as well.

Orientation-based unification models for radio-quiet AGN\footnote{Here
  we use the term `radio-quiet AGN' purely in contrast to `RLAGN' to
  indicate sources without strong radio emission or extended radio
  lobes. Few if any AGN are entirely radio-silent, and our use of this term
  does not imply a belief in a true physical dichotomy between the two
  classes; see later discussion.} \citep{Antonucci93} very successfully explain the
difference between e.g., type 1 and type 2 Seyfert galaxies in terms of
an anisotropic obscuring structure (the `torus': \citealt{Krolik+Begelman86}). This structure,
plausibly associated with the cold outer parts of the accretion flow
itself, obscures the nuclear continuum and broad emission lines in
objects where the line of sight passes through it (type 2), but allows
them to be seen directly from other lines of sight (type 1). The most
direct evidence for this picture comes from spectropolarimetry, which
reveals the broad emission lines in scattered light in type 2 objects,
showing them to be present but not directly visible to us
\citep{Antonucci+Miller85}. The simplest view of the torus as a smooth
structure is known to be incorrect, e.g., from observations of sources
where the obscuring column changes on short timescales
\citep{Risaliti+02}, which leads to the idea that different types of
AGN are selected from a distribution of both orientation and torus
covering factor \citep{Elitzur12}. However, orientation clearly has an
important remaining role to play in our view of these objects.

In the case of radio-loud objects there are two additional
complications. The first is the presence of the jet, which provides a
source of broad-band anisotropic radiation on all spatial scales. The
second is the existence of LERGs, which have no counterpart in the
Seyfert 1/2 orientation-based scheme, though they show some similarity
to the radio-quiet LINER class.

RLAGN do have one advantage, which is
that they can be selected on the basis of a roughly
orientation-independent quantity, the low-frequency luminosity
(expected to be dominated by the lobes and hence unbeamed). This means
that a low-frequency flux-limited sample such as 3CRR, is (theoretically)
unbiased with respect to orientation if redshifts can be found for all
members. \cite{Barthel89} developed the first successful
orientation-based unified model for RLAGN by noting that quasars in
the 3CRR sample in the redshift range $0.5 < z < 1.0$ had
systematically smaller projected linear sizes and brighter kpc-scale
jets and cores. He showed that an orientation of quasars within
$45^\circ$ of our line of sight was sufficient to explain the fraction
of observed quasars in the parent sample and their physical sizes.
\cite{Hardcastle+98} showed that this model could be extended to the
lower-redshift NLRGs and BLRGs in the 3CRR sample, so long as LERGs were
excluded.

At this point, the role of LERGs in unified models was unclear: for
example, it was possible that the missing narrow emission lines were
simply obscured or that the emitting material was absent, while other features of standard
AGN were still present. Work on the mid-IR and X-ray properties of the
LERGs, however, ruled this possibility out
\citep{Chiaberge+02,Whysong+Antonucci04,Hardcastle+06-2,Hardcastle+09} by showing that there was no
evidence for either heavily obscured X-ray emission or re-radiation of
obscured emission in the mid-IR, both of which are seen in NLRG. Thus
it appears that LERGs, while still possessing active jets, have no sign of a radiatively efficient accretion disk,
torus, corona, or accretion-driven emission lines, while HERG behave like
radio-quiet AGN with the addition of a jet. The nuclear optical and
X-ray emission seen from some LERGs \citep{Hardcastle+Worrall00} is consistent with coming from the
jet only. The situation is confused by the
existence of remnant sources (Section \ref{sec:remnant}), where the jet has recently switched off
--- distinguishable from active LERGs by the absence of any nuclear
emission associated with the jet --- and by a very few peculiar
objects that lack one or more of the standard AGN radiative components
\citep{RamosAlmeida+11a}, but these do not change the basic picture.

What drives the difference between LERGs and HERGs? Many authors have
proposed \citep[e.g.,][]{Ghisellini+Celotti01,Merloni+Heinz08} that
the radiative efficiency of the accretion flow is governed by the
Eddington-scaled accretion rate: only discs capable of generating more
than a few per cent of the Eddington luminosity,
\begin{equation}L_{\rm
    Edd} = \frac{4\pi GM_{\rm BH} cm_p}{\sigma_{\rm T}}
\label{eq:ledd}
\end{equation}
can produce the
optical luminosity which is directly observed in quasars and BLRG and
which drives the broad and narrow emission lines, the X-ray corona and
the mid-IR radiation from the torus. This model is supported by
observations in which the HERG/LERG classification, black
hole mass and bolometric radiative luminosity of large samples of
sources have been measured \citep{Best+Heckman12,Mingo+14} and is
consistent with expectations from theoretical disk models \citep{Rees+82,Narayan+Yi95}. In this
picture, the two classes are best referred to as radiatively
inefficient (RI: the {\it bona fide} unobscured LERGs) and radiatively efficient (RE: HERGs,
including NLRG, BLRG and radio-loud quasars). In RI objects, the
estimated jet power may greatly exceed upper limits on the nuclear
radiative luminosity.

\newcolumntype{L}[1]{>{\RaggedRight}p{#1}}
\begin{table*}
  \caption{The unified model for radio-loud sources}
  \label{tab:unif}
  \vskip 10pt
\begin{tabular}{|L{2.7cm}|L{4cm}|L{4cm}|L{4cm}|}
  \hline
  &&&\\
  &{\bf Jet at large angles to line of sight}&{\bf Intermediate
  angles}&{\bf Jet closely
  aligned to line of sight}\\
  &&&\\
  \hline
  &&&\\
  {\bf Radiatively\newline inefficient (RI)}&Low-excitation radio galaxy, LERG (FRI or
  FRII)&Low-excitation radio galaxy, LERG (FRI or FRII)&BL Lac object\\
  &&&\\
  \hline
  &&&\\
  {\bf Radiatively\newline efficient (RE)}&Narrow-line radio galaxy, NLRG (some FRI, mostly FRII)&Broad-line radio
  galaxy, BLRG, or lobe-dominated or steep-spectrum quasar (some FRI, mostly FRII)&
  Core-dominated, flat-spectrum or OVV quasar\\
  &&&\\
  \hline
\end{tabular}
\end{table*}

Claims that the RI/RE dichotomy has a one-to-one mapping to the
FRI/FRII dichotomy (that is, all RI objects are intrinsically FRIs and
vice versa, and all RE objects are intrinsically FRIIs and vice versa)
are widespread in the literature but, in their simplest form, have
been falsified by observation since 1979 \citep{Hine+Longair79}. It is
certainly the case that in the 3CRR sample almost all FRIs are RI
(with debatable exceptions such as 3C\,84), and the majority of FRIIs
are RE, but there are sufficient LERG/RI FRIIs even in that sample to
make the situation more complex. The suggestion that all of these
objects (which have nuclear X-ray radiation and VLBI-detected jets in
most cases) are simply taking a short break from being radiatively
efficient \citep{Tadhunter16} is inconsistent with observations that show
other physical differences between LERG and HERG FRIIs at
constant radio power \citep{Ineson+15}. It is also inconsistent with
our best current explanation of the FRI/FRII difference. As discussed in Section
\ref{sec:fr}, this difference is thought to come
from the interplay between the power (momentum flux) of the jet and
the extent to which it is forced to decelerate by entrainment on kpc
scales \citep{Bicknell94}. Though it is clear that the most powerful
jets will be FRII-like and the most powerful accretion flows will be
RE, there is no reason why a source which produces a jet with
kinetic power $Q$ marginally sufficient to generate an FRII-type
source with terminal hotspots in a particular environment should
necessarily also have an Eddington-scaled accretion rate high enough
to make a RE accretion flow, or vice versa. (We return to the question
of the relationship between jet power and accretion power in the
subsequent Section.)

Where does this leave unified models for RLAGN? The basic picture
remains similar to that of e.g., \cite{Urry+Padovani95} but with some
important differences in detail; in principle accretion rate, black
hole mass, jet power, obscuration covering fraction as in
\cite{Elitzur12} and angle to the line of sight are all independent
parameters of a system. The only thing that is certain is that an
object selected as `radio-loud' presumably has some non-negligible jet
kinetic power $Q$. For such a source, the accretion rate and black
hole mass determine whether the source is RI or RE, and these can vary
widely for a given $Q$ as we discuss in the next Section. The nuclear
emission from RI sources is jet-dominated at all angles to the line of
sight. For both RI and RE sources, there will be some angle to the
line of sight where the beamed small-scale jet dominates over the
optical continuum from starlight, giving a blazar-type optical
classification, but this will depend on jet power $Q$ and host galaxy
properties as well as orientation angle. Crucially, RI misaligned
RLAGN (and not, as often claimed in the literature, FRI radio
galaxies) should be the parent population of `true' BL Lac objects
with intrinsically weak lines -- this is observationally confirmed by
the existence of FRII-type structures in the extended emission from
blazars: e.g., \citealt{Rector+Stocke01}). However, the situation is
confused by objects classified as BL Lacs where intrinsically bright
lines are normally hidden by strong optical continuum
\citep{Vermeulen+95}. RE RLAGN must be the parent population of these
objects and also of flat-spectrum radio quasars. For RE sources, an
intermediate angle to the line of sight and an appropriate level of
obscuration allows a direct view of the accretion
disk/corona/broad-line region and classifies sources as BLRG or
lobe-dominated quasars; these regions do not exist in RI sources,
which are seen as LERG from all angles where the jet continuum does
not dominate. Jet power, not accretion state, and host environment
determine the large-scale radio morphology of a source. The unified
model for RLAGN discussed here is summarized in Table \ref{tab:unif}.

\subsection{Jet power and AGN power}

As yet it is poorly understood, observationally, how jet power and AGN power are
related. In RE AGN, we expect the radiative luminosity to be
proportional to the accretion rate:
\begin{equation}
  L = \eta \dot M c^2
\label{eq:radpower}
\end{equation}
where $\eta$ is the traditional efficiency factor. Here there is no
explicit dependence on black hole mass, but since accretion at
super-Eddington rates (i.e. rates that generate $L \gg L_{\rm Edd}$:
eq.\ \ref{eq:ledd})
must be short-lived, while accretion at rates much less than the
Eddington rate will give rise to a RI system, there should be a quite narrow
band of accretion rates $\dot M$ and so luminosities $L$ that can be
expected for a given $M_{\rm BH}$, scaling linearly with $M_{\rm BH}$, as observed
\citep[e.g.,][]{Steinhardt+Elvis10}.

On the other hand, jet power must be a result of a jet-generation
process and the dependences on the parameters of the system are more
complex. In the Penrose/Blandford-Znajek process
\citep{Penrose69,Blandford+Znajek77} accreting material transports magnetic field down to the event horizon
where it can be twisted by the rotation of space-time close to the
black hole; the work done in generating the jet comes directly from
the rotation of the black hole. The power that can be extracted by
this process is \citep{Tchekhovskoy+11}
\begin{equation}
  P_{\rm BZ} \approx \frac{\kappa}{\mu_0 c} \Omega_{\rm H}^2 \Phi_{\rm H}^2
\end{equation}
Here $\kappa$ is a dimensionless constant that depends on the field
geometry, and we drop a correction term that is important only for
very large spins. $\Omega_{\rm BH}$ is the angular frequency of the black hole
horizon, given by
\begin{equation}
  \Omega_{\rm H} = \frac{ac}{2r_{\rm H}}
\end{equation}
where in turn
\begin{equation}
  r_{\rm H} = \frac{GM}{c^2} \left(1+\sqrt{1-a^2}\right)
\end{equation}
and $a$ is the dimensionless black hole spin parameter
\begin{equation}
  a = \frac{Jc}{GM^2}
\end{equation}
$\Phi_{\rm H}$ is the magnetic flux threading one hemisphere of the
black hole, so loosely
\begin{equation}
  \Phi_{\rm H} = \frac{1}{2} \int \vec{B}.{\rm d}\vec{S}
  \end{equation}

The key dependences can be summarized as
\begin{equation}
  P_{\rm BZ} \propto U_B a^2 A
\end{equation}
where $U_B$ is (in some sense) the {\it ordered} magnetic field energy density at the horizon and
$A$ is the horizon area, $A \propto r_{\rm H}^2$.
It's important to
note that the field has to be ordered on large scales to give a
non-zero value of $\Phi_{\rm H}$; simply estimating the plasma $\beta$
of the accreting material is not sufficient. A dependence on black hole mass
(as $M_{\rm BH}^2$) comes from the area or radius term, so we could even more crudely
write
\begin{equation}
  P_{\rm BZ} \propto (BaM_{\rm BH})^2
  \label{eq:jp}
\end{equation}
From an observer's point of view, therefore, there are three
controlling parameters of the jet power, $Q \approx P_{\rm BZ}$. Black
hole mass can in principle be estimated from AGN properties (this is
done routinely for quasars), or estimated from the galaxy mass (or
other properties for nearby objects). Black hole spin is not
currently accessible except in the case of rapidly spinning,
radiatively efficient black holes, and is certainly not easy to
estimate for any known radio-loud AGN. And the strength of the ordered
component of field at the event horizon is completely unknown a priori
--- indeed, properties of radio AGN are the best current way we have of
estimating this quantity \citep{Zamaninasab+14}. Notice that there is
no {\it direct} dependence in eq.\ \ref{eq:jp} on the accretion rate
$\dot M$, but of course the transport of magnetic field down to the
event horizon depends on mass accretion. The key point is that the
different dependences on physical conditions at the horizon of
radiative power (eq.\ \ref{eq:radpower}) and jet power
(eq.\ \ref{eq:jp}), notably the effects of spin and the non-linear
dependence on black-hole mass in the latter, mean that we should
expect a very wide scatter in the relationship between the two
quantities both in RI\footnote{The widely cited work of \cite{Merloni+Heinz07} argues for a
correlation between the radiative output and $Q$ in {\it RI} systems,
but this is based on the use of the 2--10 keV X-ray nuclear emission as
a proxy for AGN radiative output. In the picture presented here, the 2--10 keV emission
from these systems comes from the jet itself --- see e.g.
\cite{Hardcastle+09} and references therein --- and so does not
provide any information about the AGN radiative power. The strong
correlation that they observe is essentially showing that a reasonably
constant fraction of the jet power emerges as X-rays in these systems,
which by selection are relatively unaffected by beaming.} and RE systems.

\cite{Rawlings+Saunders91} found a relationship
between the jet power $Q$ and the narrow-line luminosity $L_{\rm
  NLR}$, which is a proxy of the radiative AGN power in radiatively
efficient AGN (see above), for a small sample of powerful (3CRR) radio galaxies. Many
authors follow \citeauthor{Rawlings+Saunders91} in inferring that there is a one-to-one
relationship between accretion power and jet power. This inference is,
however, untenable in the light of what we now know about RLAGN. Three
criticisms of \citeauthor{Rawlings+Saunders91}'s work in the light of the
current picture can be made:

\begin{enumerate}
\item \citeauthor{Rawlings+Saunders91} used fairly inaccurate
  measurements of $Q$. Calculations of $Q$ from observables is
  difficult, because both ages and lobe energetics of radio galaxies
  are hard to measure (Section \ref{sec:power}). However, the main
  method they used, while it makes use of equipartition field
  strengths and spectral ages derived from those strengths and is
  therefore not correct in detail, should give a quantity that is
  proportional to $Q$. For the sake of argument, we can accept that
  this correlation really exists for the objects that
  \citeauthor{Rawlings+Saunders91} studied.

\item \citeauthor{Rawlings+Saunders91} were not aware of the
LERG/HERG dichotomy. In LERGs, since there is no significant nuclear
emission from the accretion flow, nuclear emission lines must be
photoionized either by processes irrelevant to the AGN or by the
ionizing component of the radiation from the jet itself. Since we know
that these objects contain jet-related optical and X-ray nuclear
sources even when viewed at large angles to the line of sight
\citep{Hardcastle+Worrall00-2} they should have emission-line
luminosity proportional to the luminosity of the nuclear jet, with
substantial scatter imposed by geometrical factors and the
availability of cold material in the vicinity of the nucleus. The
correlation should be quite different from that exhibited by the
HERGs, which is driven by radiative AGN power. Evidence for this can
be seen in the radio-luminosity/emission-line-luminosity plots of e.g.
\cite{Zirbel+Baum95,Hardcastle+09}. The LERGs should be excluded from
consideration in the work of \citeauthor{Rawlings+Saunders91}.
However, this does not invalidate the correlation that they observed for
high-excitation objects.

\item Crucially, \citeauthor{Rawlings+Saunders91} had a restricted sample
  composed of 3CRR objects. When much larger samples of
radiatively efficient RLAGN with AGN power indicators are considered
\citep[e.g.,][]{Punsly+Zhang11,Mingo+14,Gurkan+15} we find sources that
have much lower radio luminosities for a given accretion power than would
be expected from the correlations seen in the 3CRR objects. Thus the
central result of \citeauthor{Rawlings+Saunders91} seems to be due to
some selection bias inherent in the selection of the most powerful
radio-loud objects.
\end{enumerate}

The observations instead motivate the following straw-man model for
the relationship between $Q$ and bolometric AGN radiative power
$L_{\rm rad}$:

\begin{enumerate}
  \item Radiatively efficient AGN with a given $L_{\rm rad}$ can have jet powers $Q$ that range continuously from zero
    up to a maximum $Q \approx L_{\rm rad}$. (Note that in BZ models
    the jet power can exceed the accretion power for very high black
    hole spin parameters, but this would presumably not be the normal
    expectation.)
  \item Selection of the most radio-luminous sources selects for
    large, mature sources with the highest $Q$, for which $Q \approx
    L_{\rm rad}$: these are also the
    sources where a $L_{\rm radio}$--$Q$ correlation is expected.
    Therefore, for the most radio-luminous objects, we expect to see a
    correlation between $Q$, or radio luminosity, and $L_{\rm rad}$,
    or its proxies, as observed by \cite{Rawlings+Saunders91} and many others.
  \item However, as we relax the radio selection criteria we expect to
    see more sources that have lower jet powers (radio luminosities)
    for a given $L_{\rm rad}$, as observed \citep{Gurkan+15}
  \item In this picture, which is consistent with that of e.g.
    \cite{Kimball+11}, there is no radio-loud/radio-quiet dichotomy.
    The so-called radio-quiet quasars simply have $Q \ll L_{\rm rad}$
    (though $Q=0$ is not excluded). Below a certain jet power, radio
    emission from star formation in the host galaxy, or from other processes, may dominate the
    integrated emission \citep{Gurkan+19} though high-resolution radio
    observations may still detect a radio core.
\end{enumerate}

We conclude that observations of optical AGN properties, even setting
aside the important LERG population, are not useful
for interpreting the distribution of jet power.

\subsection{Fuelling the black hole}
\label{sec:fuelling}

From a galaxy formation perspective we can imagine two different
classes of mass to fuel black hole growth (setting aside black
hole-black hole merger): gas channelled to the central regions of the
galaxy by `secular processes' within the host galaxy itself, including the
very important channel (in massive galaxies) of cooling from the hot
phase; and gas which is brought in to the host galaxy by a merger with
a gas-rich system. The fuelling of black hole growth and hence AGN
activity connects the nature of the AGN, as discussed earlier in this
Section, with the `feedback' role of (RL)AGN in galaxy formation
models to be discussed in Section \ref{sec:feedback}.

The present authors proposed some time ago \citep{Hardcastle+07-2}
that there was a one-to-one relationship between the two fuel sources
for RLAGN (roughly speaking `hot gas' from inside the galaxy and `cold
gas' from outside) and the accretion mode (RI/RE), building on work
that suggested that the fuelling of the jets could be accomplished by
a simple Bondi flow \citep{Allen+06}. This model has now been
superseded in its simplest form for two reasons. Firstly, if we can
interpret the RI/RE difference as simply a transition in
Eddington-scaled accretion rate as discussed above, then there is no
reason why hot gas accretion should always contrive to stay below this
boundary in accretion rate or why cold gas accretion should always be
above it. Secondly, the current best understanding of accretion from
the hot phase is that it is mediated by the cooling instability
\citep[e.g.,][]{Pizzolato+Soker05,Gaspari+12,Gaspari+13}, which causes
clumps of cold material to `rain' into the centre of the galaxy; thus
the original argument that the material in a Bondi flow would be too
hot to form a radiatively efficient accretion disk is no longer
relevant or valid. It remains the case that some (mostly RE) RLAGN
appear to require a mass accretion rate far in excess of what can be
provided by cooling, and it is plausible to invoke merger-triggered
cold gas infall as a mechanism to fuel these objects. But our current view
of the relationship between the fuel source and the accretion mode in
RLAGN (set out in more detail by \citealt{Hardcastle18b}) is that there
is an association between the two rather than a one-to-one
correlation. Fuelling by (cooling-mediated) hot gas accretion will
tend to take place in massive systems with massive central black holes
and to generate rather low accretion rates, favouring RI accretion.
Fuelling through gas-rich mergers will take place in lower-mass
environments and will be capable of giving rise to high accretion
rates, favouring RE accretion. However, in this picture, it is not
safe to infer the fuel source of any particular object from its
accretion mode. These connections are of importance for galaxy feedback, and are discussed further in this context in Section~\ref{sec:feedback}.

We finally note that the {\it availability} of fuel in a particular
host galaxy should not be conflated with the actual {\it accretion} of
that fuel. It is well known that the cold gas deposited by mergers
may not trigger AGN activity until long after the merger event has
begun. Similarly, objects that are currently RI appear to be able to
accumulate quite large masses of molecular gas in the centre of the
host galaxy but are clearly not accreting it at a high rate. If this
gas represents a fuel reservoir for the AGN, it is not at all clear
how the inflow and outflow from the reservoir are controlled by
cooling. We return to this point in in Section~\ref{sec:feedback}.

\section{Host galaxies and environments}

\label{sec:hostenv}

The properties of the host galaxies in which radio-loud AGN live
provide further information with which to test our understanding of
unified models and the physical origin of the FR break
(Section~\ref{sec:pops}), but, perhaps more importantly, they allow the investigation
of crucial questions about how AGN jets are triggered, and
the energetic impact and feedback role of jets within the context of
galaxy evolution. Efforts to characterise radio-galaxy hosts date back
to the early studies of double radio sources, and the association of
Cygnus A with a galaxy merger (Section \ref{sec:opticalid}). Below we discuss current understanding of the host galaxy properties and wider environments of radio galaxies.

\subsection{Host galaxies}

A long-standing question for radio-galaxy physics, and for galaxy evolution more widely, is why some galaxies possess AGN jets and others don't: what triggers a galaxy to become radio-loud? The earliest clue to the galaxy conditions needed to trigger strong radio-loud AGN activity came from its association with massive elliptical galaxies \citep[e.g.,][]{Matthews+64}. Over the past few decades a detailed understanding of the host-galaxy properties of AGN in the local Universe has emerged, with investigations of large galaxy samples from the SDSS playing a particularly crucial role \citep[e.g.,][and references therein]{Heckman+Best14}. Below we discuss key relationships between radio AGN activity and host-galaxy properties: (1) links with host-galaxy stellar and black hole masses, (2) links with galaxy morphology and disturbance, and (3) links with star-formation properties. We also highlight any differences between observed relationships for FRI and FRII radio galaxies, and for LERG and HERG RLAGN populations. We emphasize that the majority of studies are based on low-redshift radio galaxy populations, and so some conclusions may not apply to high-redshift radio galaxy populations whose host galaxy properties remain poorly constrained.

\subsubsection{Stellar and black-hole masses}
 A strong relationship between radio-loud AGN fraction and stellar
 mass was first pointed out by \cite{Auriemma+77}, and was
 demonstrated for large samples by \cite{Best+05}, who found that more than 30 per cent of galaxies with stellar mass $M_{*} > 5 \times 10^{11}$ M$_{\odot}$ possess a radio-loud AGN with $L_{\rm 1.4GHz} > 10^{23}$ W Hz$^{-1}$. Recently, \cite{Sabater+19} have shown with more sensitive radio data that the most massive galaxies ($> 10^{11}$ M$_{\odot}$) are always `switched on' with radio AGN activity at a luminosity $L_{\rm 150MHz} > 10^{21}$ W Hz$^{-1}$ (see also \citealt{Brown+11}), and also that radio AGN fraction has a stronger dependence on stellar mass than on black hole mass. \cite{Best+05} found no significant difference in the stellar mass dependence of radio activity for optically active AGN (i.e. HERGs) and optically inactive AGN (LERGs). However, it is known that as a population, HERGs are hosted by less massive galaxies than LERGs \citep[e.g.,][]{Tasse+08,Smolcic+09,Best+Heckman12}. Morphology (FR class) has also been linked to host galaxy mass \citep[e.g.,][]{Lin+10}; however, \cite{Best+Heckman12} argue that these results are likely to be driven by the overlap between optical excitation and FR class (as discussed in Section~\ref{sec:accmodes}), with evidence that strong emission-line FRIIs (i.e. HERGs) are most distinct from FRIs (predominantly LERGs). It is worth noting, however, that a link between host-galaxy mass and radio morphology --- independent of accretion mode --- would be expected for samples matched in radio luminosity if the environmental jet disruption model for the FR break (Section~\ref{sec:fr}) is correct \citep{Mingo+19}.

\subsubsection{Mergers and interactions}
Studies over many decades have looked for signatures of disturbance in radio-galaxy hosts that could be linked to the triggering of activity \citep[e.g.,][]{Heckman+86}. There is considerable evidence for disturbance in the host galaxies of powerful radio galaxies at high redshifts \citep[e.g.,][]{Best+97}. \cite{RamosAlmeida+12} found that evidence for disturbed host-galaxy morphologies was almost universal in a sample of powerful high-excitation radio galaxies, while present at a lower level in passive galaxies in the same redshift range. Recently \cite{Pierce+19} demonstrated that the prevalence of such features in HERGs is linked to radio luminosity, with a similar but less radio luminous sample showing a lower prevalence of disturbance. \cite{Pierce+19} also note a higher prevalence of late-type galaxy hosts at lower radio powers, and suggest that secular triggering mechanisms related to disk instabilities or bars may be relevant for this HERG population. However, there is also some evidence that interaction with neighbouring galaxies is relevant for the triggering of LERGs: both \cite{Sabater+15} and \cite{Pace+Salim14} present evidence that LERG activity is influenced by interaction with neighbours, independently of large-scale environment. A recent large-sample study by \cite{Gordon+19} found a higher prevalence of major mergers in LERG hosts compared to a control population, but the overall prevalence was only 10 per cent. However, \citet{Ellison+15} found that an enhanced prevalence of LERG activity in galaxy pairs was driven by a combination of halo mass and stellar population properties, suggesting that interactions may not be directly responsible for enhanced probability of radio AGN activity in LERGs.

\subsubsection{Galaxy morphology}
It has been long established that radio-loud AGN prefer early to
late-type galaxy hosts \citep{Matthews+64}, but this does
not mean that spirals are incapable of hosting radio jets: there are now a number
of examples of spiral-hosted radio AGN with structures on $>$ kpc
scales  \citep[e.g.,][]{Ledlow+01,Croston+08,Hota+11,Mao+15,Mulcahy+16}.
Additionally, larger samples of spirals possessing smaller scale
AGN-associated radio emission have been identified by
\cite{Kaviraj+15}, who also established that the spiral hosts
typically had high stellar masses comparable to elliptical galaxy
radio AGN hosts. The observed strong relationship
between morphology and radio AGN activity may be driven partly by stellar and black hole mass differences, but is likely ultimately to be controlled by differences in galaxy evolutionary state, accretion rate and the presence of a hot gas atmosphere \citep[e.g.,][]{Krause+19}.

\subsubsection{Colours and stellar populations}
Finally, there are firm connections between radio properties and galaxy colours, stellar populations and star-formation rates. In the current picture of galaxy evolution the properties of normal galaxies  fall on a `main sequence' of star formation in which the mean star-formation rate is proportional to the galaxy mass, with a redshift-dependent normalization \citep{Elbaz+11}. At some point star formation ceases and galaxies become `red and dead', moving to a region below the main sequence in terms of star-formation rate. In the context of this picture, the vast majority of RLAGN in the local Universe are hosted by high-mass galaxies lying below the main sequence in terms of their star-formation rates \citep{Gurkan+18}. In general, radio-loud AGN are found to have lower star-formation rates than radio-quiet AGN \citep[e.g.,][]{Gurkan+15}. There are, however, differences in the star formation properties of LERG and HERG RLAGN. In addition to having lower stellar mass, HERG hosts are systematically bluer, and have higher star formation rates than those of LERGs \citep[e.g.,][]{Baldi+Capetti08,Smolcic+09,Best+Heckman12,Janssen+12,Hardcastle+13}.  Evidence has also been found for enhanced blue light in the central regions of radio-loud AGN relative to control samples \citep{Mahabal+99,Mannering+11}, suggesting enhanced star formation that could be due to simultaneous triggering of AGN and star formation activity via an inflow of gas, or due to jet-induced star formation.

\subsubsection{Summary}

In summary, it is well established that all galaxies in the local Universe are not equally capable of hosting a radio-loud AGN. Radio AGN activity is strongly linked to stellar mass --- this is also thought to be the driver of observed connections with galaxy colour/morphology, and black-hole mass. There are (at least) two possible origins for this connection, both related to the hot, hydrostatic halo typically associated with more massive galaxies at low redshifts. \citet{Krause+19} demonstrate that a substantial change in the halo density occurs at a stellar mass of $\sim 10^{11}$ M$_{\odot}$, and argue that jets of any power could be produced across the stellar mass range, but are only confined and radio-luminous in the higher-density haloes present at higher stellar mass. In reality it is likely that (at low redshifts) accretion rate is also linked to stellar mass --- if the predominantly LERG radio jets in low-redshift samples are fuelled by material cooling out of the hot halo, the inferred higher density of the hot gas halo is required to achieve sufficiently high accretion rates. 

\subsection{Large-scale environments}
\label{sec:envs}

The large-scale (galaxy group or cluster-scale) environments of radio galaxies are of interest for a number of reasons: for jets that grow to scales of tens of kpc or more they are a driving influence on subsequent jet evolution and morphology (as discussed in Section~\ref{sec:dyn}), and are the location where most of the jet energy is deposited, and for at least some radio-galaxy populations they are thought to be important for triggering and fuelling of the jets via a feedback cycle that has been particularly well studied in galaxy clusters --- we discuss this feedback cycle in Section~\ref{sec:feedback}. Environmental studies of radio jets date back more than 40 years, with optical studies indicating links between radio galaxies and Abell clusters and connections between environment and radio morphology \citep[e.g.,][]{Longair+Seldner79,Prestage+Peacock88}.

\begin{figure}
  \includegraphics[width=\linewidth]{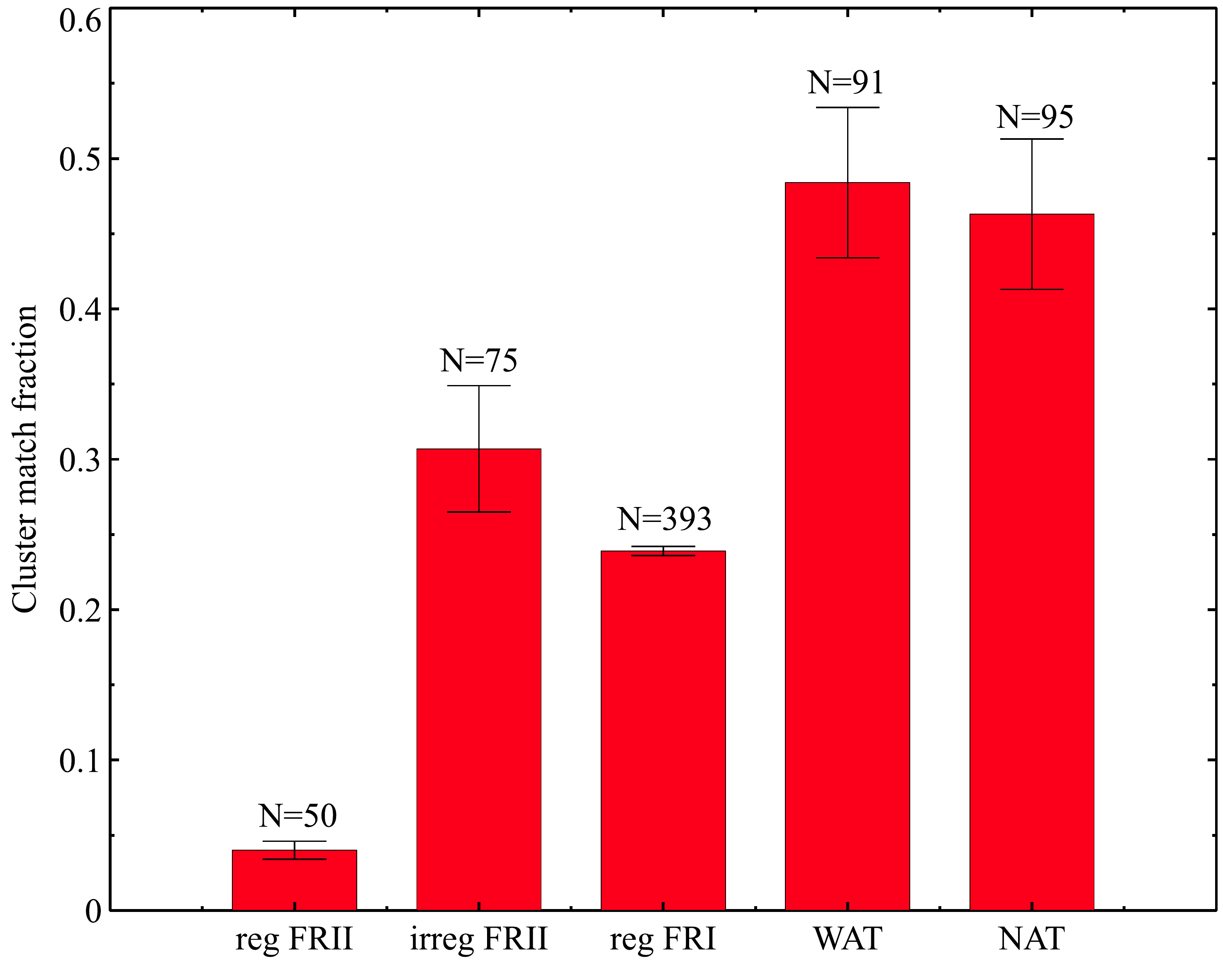}
  \caption{The relationship between radio morphology and large-scale environment for the $z<0.4$ subset of the \citet{Mingo+19} LOFAR RLAGN sample: the SDSS \citep{Wen+12} cluster match fraction and its uncertainty is determined via the method of \citet{Croston+19} for five morphological classes \citep{Mingo+19}, but with FRIIs further divided into morphologically regular and irregular subsets following the criteria of \citet{Croston+17}.}
\label{fig:lofar_envs}
\end{figure}

\subsubsection{Environments and radio properties}
The relationship between radio-galaxy properties and large-scale environment is now well determined in the local Universe. Recent studies of radio galaxies spanning a wide luminosity range in narrow redshift slices have removed the potentially confounding effect of strong flux limits in early catalogues, enabling correlations and trends with radio properties to be more carefully investigated \citep[e.g.,][]{Best04,Ineson+15}. While there are a number of famous and well-studied examples of nearby cluster-centre radio galaxies (e.g., Perseus A, M87), the bulk of the local radio galaxy population live in galaxy groups \citep[e.g.,][]{Best04,Croston+08-2,Ineson+15,Ching+17,Croston+19}. 

Radio morphology is strongly linked to large-scale environment, as shown in Fig.~\ref{fig:lofar_envs}, which shows the fraction of radio galaxies at $z<0.4$ associated with SDSS galaxy clusters ($M_{500} >  10^{14} M_{\odot}$) for different morphological classes (from the LOFAR sample of \citealt{Mingo+19}). The average cluster match fractions are significantly different for FRI and FRII radio galaxies \citep[e.g.,][]{Croston+19}, as found in many previous studies, but it is particularly striking that the morphologically regular (`classical double') FRIIs are almost never found in rich environments, unlike all other classes. 

A well-known relationship exists between bent-tailed radio galaxies and galaxy clusters: both the wide-angle tail (WATs) and narrow-angle tail radio galaxies (NATs) are found preferentially in richer environments than the general radio galaxy population \citep[e.g.,][]{O'Dea+Owen85,Mingo+19,Garon+19}. The explanation for this is thought to be the movement of the host galaxy through the intracluster medium, which leads to curving and in some case extreme bending of the jets and tails in the direction opposite to the direction of travel. It has therefore been suggested that bent radio galaxies can be used as signposts to rich environments at high-redshift \citep[e.g.,][]{Johnston-Hollitt+15,Paterno-Mahler+17}. The present authors have suggested \citep{Croston+17} that the strong preference of morphologically regular FRIIs for poor environments could also provide a powerful tool for finding and characterizing group-scale gas haloes at the epoch of cluster formation.

\subsubsection{Environments and accretion mode}
Another important conclusion from recent studies is that radio-galaxy large-scale environment is linked to accretion mode \citep[e.g.,][]{Tasse+08,Lin+10}. X-ray studies, which can provide more stringent constraints on cluster richness than galaxy number counts or two-point correlation functions, have found that low-excitation radio galaxies span the full range of environmental richness from poor groups to rich clusters, while high-excitation radio galaxies preferentially avoid rich environments \citep[e.g.,][]{Ineson+13,Ineson+15}. It is not completely trivial to disentangle LERG/HERG environmental differences from FRI/II differences, because of the strong association between FRIs and LERGs in well-studied samples, but it has been known for some time that FRII LERGs prefer rich environments more consistent with the FRI LERG population \citep{Hardcastle04}, and several studies find strong indications that accretion mode is linked to environment separately from morphology \citep{Gendre+13,Ineson+15}. For LERGs, a relationship has also been found, in both optical and X-ray environmental studies, between radio luminosity and environmental richness \citep{Ineson+15,Ching+17}, which appears not to be driven by a common link to black hole mass, and so may indicate a tight link between ICM properties and jet power. These environment--accretion mode connections provide support for AGN feedback models (see the following Section) and lend further support to arguments that accretion mode is linked to the evolutionary state of the host galaxy (Section~\ref{sec:accmodes}).

\subsection{Cosmic evolution of RLAGN, host galaxies and environments}

In the context of a galaxy evolution model where the relationship between stellar mass, accretion rate, and the evolution of a hot gas atmosphere change significantly over cosmic time, we would expect to observe considerable evolution of the properties of radio-loud AGN. Evidence for such evolution is extensive. It has been known for many decades that the space density of RLAGN was higher at early times than in the local Universe \citep[e.g.,][]{Schmidt68}. It has been shown that at lower radio luminosities, the space density begins to decline at redshifts higher than $z \sim 1$, but for higher luminosity RLAGN space density remains high out to $z \sim 3$ \citep{Rigby+11}. These changes are likely to be linked to strong differences in the evolution of low and high excitation RLAGN populations: the space density of high-excitation radio galaxies increases between $z<0.5$ and $z = 1 - 2$, while that of low-excitation radio galaxies declines \citep{Best+14,Williams+18}. The implications for these results in the context of feedback from RLAGN are discussed in Section~\ref{sec:feedback}.

We would also expect to see evolution in the host galaxy properties
for jets of a given power, but even setting aside the challenges of
making accurate jet power inferences across a wide redshift range (see
Section~\ref{sec:power}) it is not straightforward to test such
predictions due to the challenges of probing comparable ranges of
radio luminosity and rest-frame host galaxy properties at different
redshifts. The increase in space density of radio-loud AGN at $z \sim
2 - 3$ suggests that RLAGN activity must be more prevalent in lower
mass host galaxies at high redshifts, and evidence indeed supports
this conclusion. \citet{Williams+Rottgering15} found that the host galaxies of RLAGN at $1<z<2$ extend two orders of magnitude lower in stellar mass than hosts of local RLAGN \citep{Best+Heckman12} --- this population of radio galaxies in low stellar mass hosts are predominantly HERGs \citep{Williams+Rottgering15,Williams+18}. Other studies have reached somewhat different conclusions \citep[e.g.,][]{Delvecchio+17}, and deep radio surveys over wider areas (such as will shortly become available with LOFAR) should enable these questions to be investigated more fully with samples spanning a wide range in radio luminosity at high redshifts.

The conclusions relating to large-scale environments discussed in the previous Section have also been derived primarily for RLAGN
populations in the local Universe (typically $z<0.5$).  At $z>1$
environmental studies have only been possible to date for the rare objects at the high-luminosity tail of the population \citep[e.g.,][]{Hardcastle+Worrall00,Belsole+07}, leading to difficulties in making comparisons with local populations or examining relationships with radio properties. Luminous, high-redshift radio galaxies appear to be strongly associated with cluster and (at $z>2$) protocluster environments, and have proved a useful tracer of the richest overdensities at $z>2$ \citep[e.g.,][]{Venemans+07,Miley+DeBreuck08,Wylezalek+13,Hatch+14}. With new and upcoming surveys, it will be possible to obtain a more complete picture of high-redshift radio galaxies across the full radio luminosity range. Simple hydrodynamical considerations suggest that radio galaxies of similar luminosity and size (and hence internal pressure) at low and high redshift must be embedded in gas at similar pressures. For this reason the morphology and size distributions of high-redshift radio galaxy populations as a function of luminosity will also provide insights into how the environments of radio galaxies evolve with redshift. 

\section{The feedback role(s) of RLAGN}

\label{sec:feedback}

\begin{figure*}
  \includegraphics[width=\linewidth]{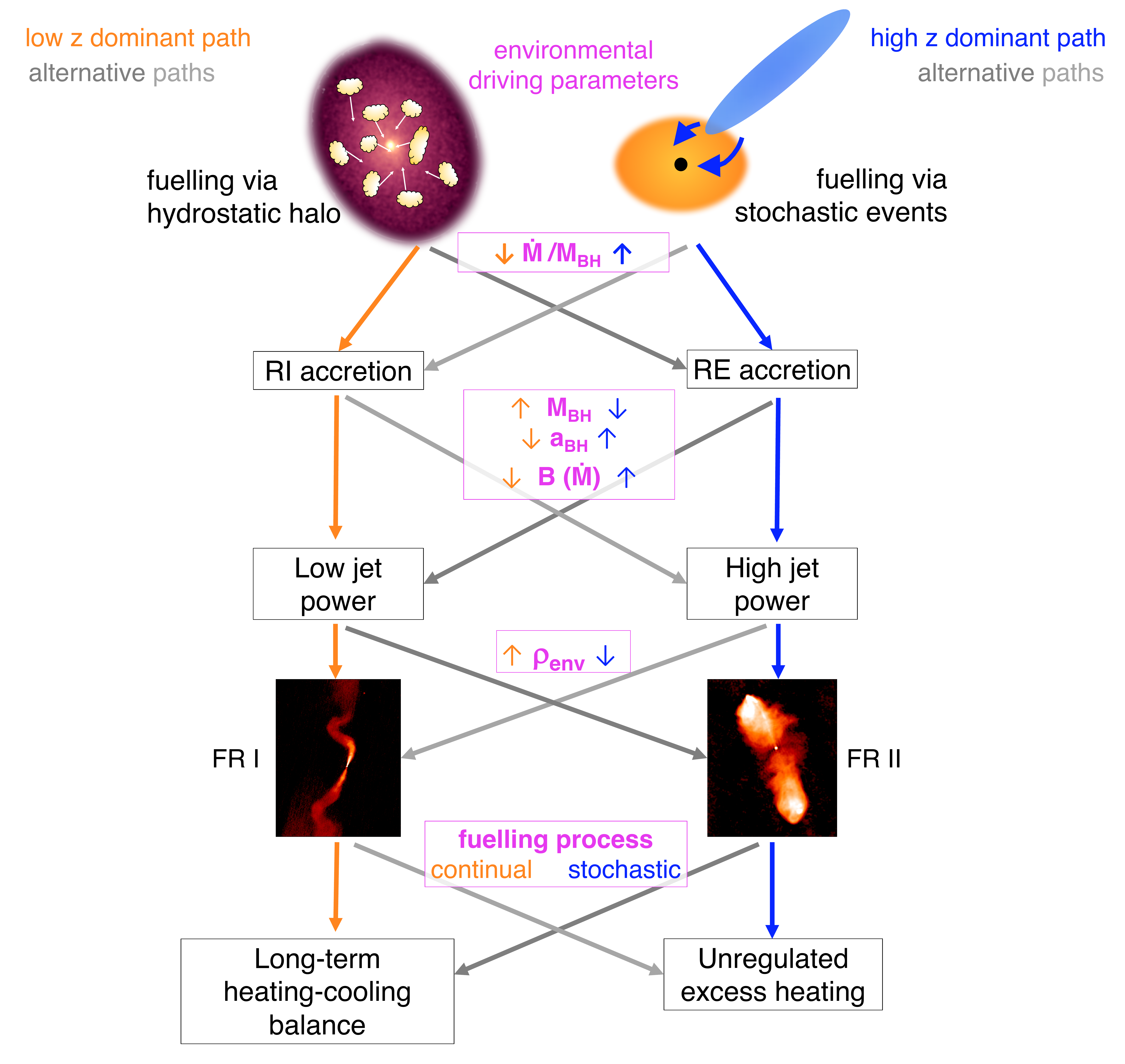}
  \caption{A diagrammatic summary of the relationship between fuelling, accretion, jet production and feedback, as described in Sections~\ref{sec:accmodes} and~\ref{sec:feedback}. In principle systems can follow any of the arrows connecting accretion and jet behaviour, but the choices of pathway are controlled by a series of environmental parameters (magenta boxes), which are interconnected and closely linked to the galaxy's evolutionary history.}
\label{fig:accfb}
\end{figure*}

Much recent research on radio-galaxy populations has been motivated
--- at least in part --- by the now widespread acceptance that RLAGN
are an important galaxy feedback mechanism, playing a key role in the
evolution of galaxies and large-scale structure
\citep[e.g.,][]{Cattaneo+09}. Wider discussions of AGN feedback in
galaxy formation models can be found in the recent reviews of
\citet{Somerville+Dave15,Naab+Ostriker17}, and of the observational
evidence for AGN feedback in the reviews of \cite{Fabian12} and \cite{McNamara+Nulsen07, McNamara+Nulsen12}.  In this section we summarize current understanding of the potential galaxy feedback role(s) of RLAGN as indicated by cosmological simulations and galaxy evolution models, before discussing observational evidence of jet impact and the feedback roles of different radio-galaxy sub-populations. Fig.~\ref{fig:accfb} draws together the discussions of RLAGN populations, accretion and jet power earlier in the Chapter with the authors' perspective on the potential feedback roles of AGN jets, which we explain further in the sections that follow. 

\subsection{The need for AGN jet-driven feedback in galaxy evolution}
Galaxy feedback processes that regulate the formation of stars and
growth in stellar mass are a key element of modern galaxy formation
models \citep[e.g.,][]{Somerville+Dave15,Naab+Ostriker17}. A wide range
of processes associated with star formation and black-hole growth
affect the thermodynamical, chemical and star formation histories of
galaxies, and much observational and theoretical/computational effort
is currently being devoted to disentangling and quantifying these
processes. The general argument that feedback processes are important
is not controversial, but there is considerable debate about the
extent to which some form of AGN feedback is needed to explain
particular mismatches between models and observations, about the
microphysics of feedback mechanisms and energy transport within the
ISM and ICM, and about the relative importance of different AGN
feedback mechanisms (which include the effects of winds, jets, cosmic
rays, and radiation). Feedback from radio-galaxy jets is in some ways
one of the best understood aspects of this problem, and we focus
specifically on the potential role of mechanical feedback from AGN
jets in the discussion that follows. A wide range of observational
constraints on jet mechanical feedback in the local Universe have been
assembled over the past two decades, particularly from {\it Chandra}
and {\it XMM-Newton} observations of galaxy clusters and groups, and
more recently from {\it Hitomi} and ALMA --- these will be discussed
in the next section. However, many questions still remain, including
the details of how, when and where a feedback loop operates and the mechanisms by which the jet energy is coupled to surrounding gas in different contexts, as well as broader questions about the role of jets in different environments and at different epochs. 

What problems can jet kinetic feedback solve? AGN feedback considered
more broadly has been invoked in models of galaxy evolution primarily
to address two major mismatches with observations: (i) the need to
suppress star formation in the most massive galaxies in order to
reproduce the high-mass end of the local galaxy luminosity function
\citep[e.g.,][]{Benson+03} and prevent cooling flows in galaxy
clusters \citep[e.g.,][]{Peterson+03,Sakelliou+02}, and (ii) the need
to explain the origin of the strong colour bimodality of galaxies \citep[e.g.,][]{Strateva+01,Baldry+04}, which requires rapid quenching of star formation, moving star-forming galaxies to the red sequence of quiescent galaxies. AGN feedback is also implicated in the origins of the well-known galaxy black-hole -- bulge mass correlation, as suggested by \citet{Silk+Rees98}, although such a relation can arise through mergers in the absence of feedback self-regulation \citep{Peng07}. Finally, the evolution of the gas mass fraction of dark matter haloes, the properties of the circumgalactic medium, and the thermodynamic properties of gas in galaxy groups and clusters, are all highly sensitive to the injection of energy from AGN over a wide range in redshift \citep[e.g.,][]{McCarthy+11,LeBrun+14,Davies+19,Voit+18,Kauffmann+19}.

Modern hydrodynamical cosmological simulations take a range of
approaches to modelling black hole growth and consequent AGN feedback,
ranging from a single mode of feedback controlled by the mass
accretion rate, coupling to the ISM with fixed efficiency
\citep[e.g.,][]{Schaye+15,McCarthy+17}, to different modes associated
with high and low accretion rates that may or may not explicitly be
linked to specific outflow types or coupling mechanisms
\citep[e.g.,][]{Vogelsberger+14,Weinberger+18,Dave+19}. The majority of
implementations are in practice agnostic about the relative contributions of
AGN jets, winds and radiative feedback, but provide insights
into the importance of different accretion rate regimes in influencing
particular aspects of galaxy growth \citep[e.g.,][]{Rosas-Guevara+16}.
In general it is found that most black hole growth for moderately
massive black holes must take place when AGN are accreting at a high
fraction of the Eddington rate, so that radiatively efficient AGN
activity is associated with the evolution of the $M_{\rm BH}$--$M_{\rm bulge}$ relation. Low Eddington-scaled accretion rates occur at later times, with the associated radiatively inefficient AGN activity linked to the suppression of star formation in massive galaxies. There is less consensus as to which accretion regime is relevant for the quenching of star formation by AGN \citep[e.g.,][]{Bower+17,Terrazas+19}.

The association of radiatively inefficient accretion with jet generation (see also the X-ray binary context elsewhere in this volume), together with the strong observational evidence for the impact of jets in massive galaxies at low redshift, initially led to a widely discussed paradigm of separate `radio' and `quasar' modes of AGN feedback, with the former responsible for solving the mismatch at the massive end of the galaxy luminosity function, and the latter associated with black-hole growth \citep[e.g.,][]{Croton+06}. There remains a focus in the literature on jet kinetic feedback as a `maintenance' process that regulates star formation at late stages of evolution \citep[e.g.,][]{Fabian12} --- we discuss this scenario further in the next two subsections. However, as we emphasise in Section~\ref{sec:accmodes} and in Fig.~\ref{fig:accfb}, powerful AGN jets also occur at high Eddington-scaled accretion rates, with increasing prevalence towards the peak of cosmic star formation, and so in Section~\ref{sec:highz} we also consider the potential relevance of feedback from RLAGN for other aspects of galaxy formation including at earlier stages in galaxy evolution. 

\subsection{Observational evidence for jet kinetic feedback}

The observational evidence for the energetic impact of RLAGN jets in
the local Universe is extensive and we only summarize it briefly here --- see \citet{McNamara+Nulsen07,McNamara+Nulsen12} for more comprehensive discussion. Direct estimates of jet energy input come from X-ray surface brightness deficits in galaxy clusters and groups (`cavities'), from detections of shock fronts associated with expanding radio lobes on galaxy scales and in clusters, and from ripples believed to be transporting and spreading out the injected energy through the ICM gas in cluster cores. Indirect energy budget estimates based on improved knowledge of the physical conditions within radio-galaxy lobes provide corroborating information, as do increasingly sophisticated hydrodynamical simulations of radio-lobe propagation that reproduce observations. 

Cavities excavated by expanding radio lobes were first identified in
{\it ROSAT} images \citep{Bohringer+93,Hardcastle+98} and subsequently found in many clusters and studied in depth using {\it Chandra} \citep[e.g.,][]{Dunn+Fabian04,Dunn+05,Birzan+04,Birzan+08}. They provide a direct measure of the $p$d$V$ work carried out on the surrounding medium as radio lobes expand (Section \ref{sec:power}), and such observations have demonstrated firmly that sufficient energy is being transferred from expanding radio galaxy lobes to offset the current rate of gas cooling in the centres of cool core clusters in the nearby Universe. Cluster cavities have been detected out to $z\sim1$ \citep{Hlavacek-Larrondo+15}, but there remain observational limitations and selection effects that affect our ability to draw robust population-wide conclusions \citep[e.g.,][]{Birzan+12}. There is also substantial observational evidence for energy transfer from jets to their environments via shocks, which have been detected in a range of environments \citep[e.g.,][]{Kraft+03,Croston+07,Forman+07,Croston+09,Randall+15}, on scales ranging from the central ISM to hundreds of kpc.

While shock heating is likely to be important in some situations, the
identification of ripple features in the intracluster medium of the
Perseus cluster \citep{Fabian+03,Fabian+06} demonstrated that there is
a means of distributing jet energy injection azimuthally to influence
the entire cluster core region in which heating is needed to balance
cooling. Similar ripples have now been identified in several other
clusters and groups \citep{Forman+05,Sanders+Fabian08,Blanton+11}.
Recently the (sadly short-lived) {\it Hitomi} mission was able to make the first precise measurements of gas motions in the core of the Perseus cluster \citep{hitomi16}, providing constraints on the energy stored in turbulent motions and the nature of energy transport. There has been some debate about the apparent `quiescent' nature of the gas in the cluster core region revealed by {\it Hitomi}, but several studies have shown that this is not inconsistent with the gentle mode of AGN feedback heating that appears to be in operation in cool-core regions \citep[e.g.][]{lau+17,fabian+17}.

Further important observational clues to the nature of jet feedback come from a wealth of observations of atomic and molecular filaments and outflows of gas in the central regions of galaxies hosting RLAGN. Evidence for outflowing atomic gas has been found in both powerful RLAGN \citep[e.g.,][]{Tadhunter91} and `radio-quiet' systems with small-scale jets \citep[e.g.,][]{Morganti+98,Rupke+Veilleux11} --- see \citet{Morganti+Oosterloo18} for a recent overview of outflow properties inferred from HI absorption studies. There is growing evidence, particular from recent ALMA studies, of massive outflows of molecular material entrained or uplifted by jets or rising radio lobes \citep[e.g.,][]{Alatalo+11,Dasyra+15,McNamara+14,Russell+14,Russell+16,Russell+17,Tremblay+18}. A number of these examples are cool-core clusters, the environments in which AGN feedback is required to act most strongly to suppress cooling and star formation. These systems have long been known to contain spectacular filamentary nebulae \citep[e.g.,][]{Crawford+99b}, of both atomic and molecular gas \citep[e.g.,][]{Edge01,Hatch+05,O'Dea+08}. The physics of these filaments is complex, and their origins are still under debate, but substantial evidence points to cold, low entropy gas being lifted from the cluster centre to distances of tens of kpc, most likely in the wake of rising radio bubbles \citep[e.g.,][]{Fabian12}. 
The kinematics and locations of cold gas in cooling hot gas haloes
provide crucial clues to how jet feedback from AGN can self-regulate
so as to maintain a long-term balance between heating and cooling as
required by observations \citep[e.g.,][]{McDonald+18} and by galaxy
evolution models (see previous Section).

\subsection{Self-regulation and heating-cooling balance}

In parallel with
observational advances in studying heating and cooling in hot
hydrostatic haloes, there has been much research around the mechanisms
of achieving a self-regulating AGN feedback loop. As mentioned in
Section \ref{sec:fuelling}, it has been proposed
that thermally unstable gas cooling is triggered under conditions
related to the ratio of cooling to free-fall time
\citep[e.g.,][]{Sharma+12,McCourt+12,Gaspari+12,Voit+15}, leading to
the condensation of clumps of cold material that `rain' onto the
central AGN, losing angular momentum via collisions so as to accrete
onto the central black hole
\citep[e.g.,][]{Pizzolato+Soker05,Gaspari+12,Gaspari+13}. This
`chaotic cold accretion' (CCA) powers the RLAGN, leading to outward flow of energy
and consequent heating, and so enabling self-regulation of the cooling process. Observations of gas conditions in galaxy, group and cluster halos \citep[e.g.,][]{McNamara+16,Hogan+17,Pulido+18} suggest that the uplift and movement of thermally unstable gas driven by the AGN outflow may play a crucial role in stimulating the self-regulating feedback cycle. A more in-depth discussion of current debates around the physics of heating, cooling and the AGN feedback loop in hot atmospheres can be found in the recent review of \citet{Werner+19}.

As well as detailed individual and small sample studies of jet
feedback, there have been a number of attempts to assess the
population-wide balance between cooling and heating in hot atmospheres
(on galaxy, group and cluster scales). Such estimates rely on a
well-determined radio luminosity function (and ideally also well
constrained evolution of the luminosity function), and on robust
methods to translate from radio luminosity to jet kinetic power. The
radio luminosity function is now well determined in the local Universe
\citep{Mauch+Sadler07}. There remain substantial caveats in converting
to jet power (see Section~\ref{sec:power}), but it is possible to draw
some reliable general conclusions about the heating and cooling
balance at low redshifts. \citet{Best+06} compared the rate of heating
via mechanical jet energy input, based on an estimate of the local
radio luminosity function obtained from cross-correlation of SDSS with
NVSS and FIRST and conversion to jet power via X-ray cavity relations,
to cooling rates in elliptical galaxies determined from the
relationship between X-ray and optical luminosity, finding remarkably
good agreement. \citet{Smolcic+17} recently used the deep COSMOS field
radio data to investigate the cosmic evolution of the RLAGN `kinetic
luminosity function' finding good agreement with the results of
\citet{Best+06} at low redshift, with the kinetic luminosity density
increasing out to $z \sim 1.5$ and then declining gradually towards $z
\sim 5$. Observational constraints on cooling rates do not exist
beyond $z \sim 1.5$, but they find reasonable agreement with the
cooling rates required in the semi-analytical model of
\citet{Croton+16}. The present authors recently constructed
\citep{Hardcastle+19} a sample of $\sim 23,000$ RLAGN from the LoTSS
DR1 catalogue and used a new analytic model for radio-lobe dynamical
evolution, accounting for the effects of environmental variation,
radiative losses and redshift, to obtain more realistic conversions
from radio luminosity to jet power and thus make similar estimates of the kinetic luminosity function and overall heating rate at $z<0.7$. As with previous work, it is concluded that the rate of heating from RLAGN jets in the local Universe is well matched to the cooling luminosities of galaxy groups and clusters. 

At the level of individual objects it remains unclear how tightly cooling and heating processes are coupled across the full range of RLAGN environments at low redshift. The observational evidence in support of the CCA mechanism relates mainly to brightest cluster galaxies, while, as noted in Section~\ref{sec:accmodes}, a number of nearby radio galaxies possess apparently stable disks of molecular gas \citep[e.g.][]{Lim+00,Prandoni+10} whose origin and relationship to AGN fuelling is unclear. Nevertheless, while the observational picture remains complex and timescales relating to heating and cooling balance poorly constrained, there is now substantial evidence for a self-regulated feedback scenario in massive systems at low redshift, linked to the presence of hot hydrostatic haloes. We suggest that the pathway indicated by orange arrows in Fig.~\ref{fig:accfb} represents the most common RLAGN population at low redshifts: RI accretion ultimately originates from cooling out of a hot-gas halo, with long-term balance between heating and cooling mediated by low-power, FRI morphology jets. However, as indicated by the grey arrows, the RLAGN population is complex, and under some conditions RE sources (HERGs) and/or FRIIs will also participate in self-regulated feedback. Conversely, not all RI and/or FRI morphology sources will be in environments where heating and cooling are in balance.

\subsection{Jet kinetic feedback at high redshifts}
\label{sec:highz}

It is difficult to extrapolate the self-regulated RLAGN feedback scenario described in the previous Section to higher redshifts. One reason is our limited knowledge of the high-redshift radio luminosity function, and another is the increasingly large systematic uncertainty and potential biases in conversions from radio luminosity to jet power beyond the local Universe, due to the increasing importance of radiative losses and increased uncertainty in environmental properties (see Section~\ref{sec:cond}). A further important factor is redshift evolution in the distribution of accretion mode for RLAGN. Section~\ref{sec:accmodes} discussed our current understanding of accretion mode in RLAGN: the local population is dominated by radiatively inefficient (RI) systems, but a substantial population of radiatively efficient (RE) RLAGN exist, and are more prevalent at higher redshifts, as discussed in the previous Section. This evolution in the RLAGN population may have interesting implications for jet feedback at $z>1$.

As argued above (Sections \ref{sec:accmodes}, \ref{sec:fuelling}), accretion mode in RLAGN is controlled by the ratio of accretion rate to black-hole mass, and not by the source of accreting material: in principle the chaotic cold accretion mechanism discussed in the previous section can achieve accretion rates high enough to power RE RLAGN, particularly in systems with lower mass black holes. However, powerful RE (high-excitation) RLAGN are observed to be located systematically in lower mass galaxies and poorer large-scale environments than RI RLAGN of similar inferred jet powers (Section~\ref{sec:hostenv}). From the perspective of self-regulating feedback, this population appears anomalous: their accretion rates (in absolute terms) from CCA, or other processes related to the hot-gas halo, must be lower than for their RI counterparts in richer haloes and more massive host galaxies, but their jets are transporting similar amounts of energy into their surroundings. The simplest explanation --- consistent with the high prevalence of galaxy merger signatures in the hosts of the most powerful high-excitation RLAGN \citep[e.g.,][]{RamosAlmeida+12} --- is that powerful RE systems achieve the high accretion rates necessary to power their jets via an additional mechanism of cold gas inflow driven by galaxy mergers and interactions. The energetic output of powerful RE systems is then decoupled from a self-regulating feedback loop, so that we might expect larger imbalances between heating and cooling in these systems compared to those (predominantly, but probably not exclusively, RI) systems that are accreting only from their hot-gas halo. We note that the conclusion that powerful HERGs inhabit poor environments derives from observations at $z<1$ (Section~\ref{sec:hostenv}), but simple hydrodynamical arguments indicate that radio galaxies of similar jet power and size will inhabit similarly rich large-scale gas haloes at any redshift \citep[e.g.,][]{Croston+17}. We suggest that a substantial proportion of the $z>1$ HERG population are likely to be `over-heating' their environments: this population could be responsible for the known excess entropy present in hot gas haloes at low mass \citep[e.g.,][]{Pratt+09,Short+10,Fabian12}. We indicate this feedback mode as the possible endpoint of the dominant high redshift pathway, indicated by blue arrows, in Fig.~\ref{fig:accfb}, but again emphasise that under appropriate conditions sources may instead follow the grey pathways in which accretion, jet and feedback properties deviate from the majority behaviour. Such populations could form significant sub-populations for particular combinations of redshift and galaxy/halo mass.

Another high-redshift jet population of particular interest are the
galaxy-scale jet structures found in `radio-quiet' quasars
\citep{Jarvis+19}, Seyferts \citep{Gallimore+06,Morganti+99,Mingo+11},
and in ordinary galaxies at low redshift
\citep[e.g.,][]{Croston+07,Croston+08}, as discussed in
Section~\ref{sec:pops}. The prevalence of these radio outflows on
scales of a few to several tens of kpc is not yet well determined
either in the local Universe or during the epoch of peak quasar
activity, although high-$z$ examples of jets interacting with the IGM
are known to exist \citep{Nesvadba+17}. Kinetic feedback from small jets on galaxy scales could therefore comprise an overlooked feedback mechanism during the epoch of high accretion rates and black hole growth, and exciting opportunities to investigate this question will be provided by upcoming sensitive, high resolution radio facilities. Better constraints on the energetic impact of the variety of jet sub-populations expected to be present at the peak of quasar activity and black-hole growth will make it possible to quantify the relative contributions of winds and jets at this epoch, and will inform substantial improvements to feedback treatments in cosmological simulations. 

More generally, by obtaining well-determined luminosity functions at $z\sim2$ and beyond, down to luminosities corresponding to the dominant populations in well-studied local samples \citep[e.g.,][]{Best+Heckman12}, surveys such as those with LOFAR and MeerKAT, should lead to the first robust estimates of the energy available from radio jet feedback at the peak of star formation and quasar activity, and enable the host-galaxy and large-scale environmental properties of these new populations of lower luminosity high-$z$ RLAGN to be determined. Complementary constraints on the evolution of baryons in the presence of jet and wind feedback from AGN will come from future, more sensitive X-ray facilities, such as {\it Athena} \citep{Nandra+13}, that will directly measure group-scale hot-gas atmospheres at $z>2$, and trace the evolution of group and cluster gas entropy profiles. We therefore look forward to an improved understanding of the relevance of jet kinetic feedback beyond the local Universe over the next decade.

\section{Astrophysical uses of radio galaxies}

\label{sec:uses}

In this Section we briefly discuss the relevance and use of RLAGN for other areas of astrophysics, namely measurements of cosmic magnetic fields, the non-thermal content of galaxy clusters, cosmology, and cosmic rays.

\subsection{Cosmic magnetism}

The origins and evolution of magnetic fields in the Universe are a substantial uncertainty in structure formation models, and an important science driver for the Square Kilometre Array. Faraday rotation techniques (see Section~\ref{sec:obs}) have been used to measure magnetic field strengths and structure in a range of astrophysical environments. 

Magnetic field strengths in groups and clusters of galaxies can be measured both via embedded radio galaxies, and radio galaxies located beyond the group/cluster but along a line of sight that passes through the group/cluster gas. Faraday rotation studies of cluster-centre radio galaxies date back to the 1970s, with radio galaxies at the centre of rich, `cooling flow' clusters found to have high rotation measures \citep{Carilli+Taylor02}. Studies of background radio galaxies have also been used to measure cluster magnetic field strengths \citep[e.g.,][]{Clarke+92}. More recently this approach has enabled mapping of cluster magnetic field distributions \citep{Bonafede+10} and investigation of relationships between cluster thermodynamic conditions and magnetic fields \citep{Govoni+10}. Cluster and group magnetic field distributions provide important constraints on ICM transport processes and on models for the origin and evolution of their magnetic fields.

A related topic of interest is the potential role of AGN in injecting magnetic fields \citep[e.g.][]{Xu+11}, and/or altering the magnetic field structures within galaxy groups and clusters. Recent high-resolution rotation measure studies of resolved radio galaxies in galaxy groups with well-measured gas density distributions have revealed complex magnetic field structure, with ordered field components associated with compression of the gas as the radio lobes expand \citep{Guidetti+11,Guidetti+12}. Detailed studies with current and future radio instruments should enable further advances in understanding the influence of radio galaxies on magnetic field properties of groups and clusters.

Radio galaxies also have the potential to be used to provide `rotation measure grids' --- sufficiently many strongly polarized radio sources across the sky will enable magnetic fields to be determined on a range of scales and cosmic environments \citep[e.g.,][]{Beck+Gaensler04,Johnston-Hollitt+04,Krause+09}. With SKA pathfinders and eventually the SKA itself it should be possible to build grids of background sources that will enable the magnetic field of the Milky Way to be mapped on arcmin scales, as well as enabling detailed investigations of magnetic fields in nearby galaxies via the effect of propagation of emission from background radio galaxies through their ISMs. A recent LOFAR study has demonstrated the potential of this technique for studying filaments of large-scale structure \citep{O'Sullivan+19}. However, even with sensitive polarimetry over broad frequency ranges there remain challenges in disentangling multiple contributions to the observed rotation measure, including material intrinsic to the background sources being used. Nevertheless, over the next decade, radio galaxies are likely to be a powerful tool for understanding magnetic fields in a range of environments.

\subsection{Non-thermal particle populations in clusters}

Galaxy clusters contain an important non-thermal particle population, which contributes to pressure support within the cluster, and results in diffuse extended radio sources, known as radio haloes and relics. Halos are Mpc-scale structures, thought to be caused by turbulent reacceleration of particles pervading the ICM. Relics are narrow features, sometimes found in pairs \citep[e.g.,][]{VanWeeren+11}, thought to trace shock waves in the ICM. For a detailed review of these diffuse cluster radio sources, see \citet{feretti+12}. 

Particle acceleration models for diffuse radio sources in clusters
suggest that a seed population of relativistic particles is required
to produce the observed extended radio structures
\citep[e.g.,][]{Brunetti+Jones14}. RLAGN are an obvious source
for such a particle population, and low-frequency radio observations
are beginning to provide more concrete evidence in favour of this
picture. Several examples of radio `phoenices', relic structures
whose morphology and/or spectral structure indicate a
revived/re-accelerated region of plasma associated with an AGN, have
recently emerged \citep[e.g.,][]{Bonafede+14,VanWeeren+17,VanWeeren+19}, as well as examples of very extended and complex radio-galaxy tail structures in clusters \citep{Hardcastle+19b,Clarke+19}. 

Hence, it appears that there are strong links between remnant radio lobes and non-thermal particle populations in galaxy clusters. Low-frequency studies, and particularly broad-band spectral investigations, should enable substantial progress in determining the long-term effects of RLAGN on the properties of the intracluster plasma. Additionally, the mixing of radio-lobe plasma into the ICM is expected to be important for the evolution of cluster magnetic fields \citep[e.g.,][]{Xu+10}. Better constraints on these processes will enable improved modelling of energy transport processes within the ICM and feedback effects on the evolution of cluster baryon content.

\subsection{Radio galaxies in cosmology}

Historically RLAGN (including quasars) were used as signposts of
high-redshift objects; as discussed above, early identifications of
RLAGN with quasar hosts opened up the study of the high-redshift
universe. Throughout the 80s and 90s, the practice of selecting
candidate high-redshift galaxies starting from radio surveys led to
samples of hundreds of objects with $z>2$ being compiled
\citep{McCarthy93,Miley+DeBreuck08}. More recently, of course,
high-$z$ objects can be more conveniently selected from large-area,
deep optical sky surveys and radio selection is no longer a primary
tool. Nevertheless, interest in finding high-redshift, powerful
objects persists: if even a single bright source could be found at a
redshift $z>6$, in the presumed Epoch of Reionization, then redshifted
21-cm absorption against its synchrotron continuum, the so-called
21-cm forest, would provide a unique probe of the state of matter at
that point in the early universe \citep{Carilli+02b}. Currently the
redshift record holders are just below $z=6$
\citep{Banados+18,Saxena+18} and it is not clear whether powerful
radio galaxies can even exist at much higher redshifts, given the
effects of the CMB on inverse-Compton losses in the lobes. Future
radio surveys will enable much deeper searches for high-$z$ objects.

RLAGN have various other applications as tracers of large-scale
structure when cross-matched with the CMB or with optical surveys (see,
e.g., \citealt{Raccanelli+12}), constraining models of dark energy
and/or modified gravity. However, these techniques require large-area,
homogeneous survey data that do not yet exist for their full
effectiveness. One cosmological application that has already been
explored is the use of radio galaxies as standardizable rulers
\citep{Daly94,Daly+Guerra02}. This approach, which has a long history
\citep[e.g.,][]{Hoyle59,Kapahi87} is somewhat similar to the
use of supernovae as standardizable candles, but has the disadvantage
that it relies on a particular model of radio source evolution when
the environments of radio sources, particularly at high $z$, are not
well understood. 

\subsection{Radio galaxies and the origin of cosmic rays}
\label{sec:uhecr}

It has long been clear \citep{Hillas84} that the large volumes and
strong magnetic fields in radio galaxies mean that their large-scale
components (lobes and hotspots) are possible sites of the acceleration
of the highest-energy cosmic rays, with energies above $\sim 10^{19}$
eV (hereafter ultra-high-energy cosmic rays, UHECR). Observationally
the presence of high-energy leptons (albeit at much lower energies)
implies that efficient hadronic particle acceleration is possible, and
inverse-Compton measurements allow us to estimate the field strengths
in the acceleration regions. Additional constraints are that the
sources of UHECR must be nearby, since UHECR suffer from strong
attenuation due to photopion production on the cosmic microwave
background and/or photodisintegration of nuclei on scales of the
Greisen-Zat'sepin-Kuzmin cutoff (GZK: \citealt{Greisen66}) of $\sim
100$ Mpc, and that they must be capable of accelerating not just
protons but also heavy nuclei, since a heavy nuclear component seems
to be necessary to explain the composition observations
\citep{Taylor14}. The first constraint disfavours FRII hotspots as the
dominant sources of UHECR, since their space density is very low and
there are few within the GZK cutoff; the dominant population is
low-power sources and many such systems with the capability to confine
UHECR exist within 100 Mpc \citep{Hardcastle10}. The second constraint
requires nuclei to be found inside radio galaxy lobes, but entrainment
of stellar winds can permit that \citep{Wykes+15}.

The acceleration {\it mechanisms} for UHECR in these sources are less
clear, and we defer detailed discussion to another review in this collection. Possibilities
include first- or second-order Fermi acceleration in the lobes
\citep{Hardcastle+09,Matthews+18}, acceleration in the jets
\citep{Honda09,Meli+Biermann13} or some combination of the two. As yet
there is no model that relates the distribution of the
size/luminosity/jet power of RLAGN to the observed cosmic ray flux,
sky distribution, spectrum and energy-dependent composition measured
at Earth, but many of the ingredients for constructing such a model
now exist. The discovery of a high-energy neutrino plausibly
associated with a flare in the jet of the blazar TXS 0506+056
\citep{IceCube+18} should give new impetus to the development of such
models.

\section{Future prospects}

\label{sec:future}

A great deal has been learnt about radio galaxies and other RLAGN in
the hundred years since the first observation of non-thermal radiation
from a jet \citep{Curtis18}. We start the next century of RLAGN
studies with the clear idea that these objects, almost invisible in
traditional optical studies, have a profound effect on their host
environments and on the evolution of galaxies in the Universe. We have
also developed a good working understanding of the basic dynamics of,
and physical conditions in, the large-scale structures that are the
main focus of this review.

Many challenges remain. Observationally, progress in the radio is
expected to be rapid as a result of the next generation of radio
instruments, particularly in the realm of radio surveys: wide and deep
radio surveys with LOFAR, ASKAP, MeerKAT and the forthcoming SKA are
starting to allow radio astronomers to catch up with optical astronomy
in terms of sheer numbers of sources. Surveys now being carried out
have the capability to detect all objects where radio AGN activity
dominates over star formation out to very high redshifts.
High-fidelity, high-resolution imaging is still a little way behind,
but long-baseline LOFAR and the mid-frequency SKA both have the
capability to provide this. Time-domain radio work and broad-band
polarimetry are likely to be other fruitful areas in the coming years.
In the optical, wide-area radio surveys rely on the next generation of
wide-area optical surveys for optical identification --- this includes
the existing PanSTARRS and Legacy surveys in the Northern hemisphere
and the forthcoming LSST surveys in the South. Complementary
spectroscopic surveys, or good photometric redshifts, are also needed
to make progress. In the X-ray, the other key area for RLAGN studies
because of its capability to probe environments, magnetic fields and
particle acceleration as well as to allow the study of (radiatively
efficient) AGN activity, we can expect interesting results on host
environments in the nearby universe, from the recently launched {\it
  e-ROSITA}, while in around a decade's time {\it Athena} should
provide the capability to carry out very detailed studies of the
feedback mechanisms and the dynamics of the hot gas, as well as the evolution of RLAGN environments. A key complement
to {\it Athena} would be a high-resolution, high-sensitivity X-ray
telescope --- the US mission concepts {\it Lynx} and/or {\it AXIS}
could provide this capability. Again, time-domain studies in the X-ray
have great potential for studies of the particle acceleration
mechanism in jets and hotspots, and are so far only possible in a
handful of bright nearby objects. In high-energy gamma rays, the Cerenkov
Telescope Array (CTA) will allow the resolution of a number of nearby
objects, as discussed in Section \ref{sec:gamma}, and should give
inverse-Compton constraints on magnetic fields for a number of FRI
sources that are currently inaccessible to inverse-Compton studies, as
well as potentially constraining particle content through the
detection of accelerated high-energy protons.

In terms of modelling, we expect to see the gradual convergence of
cosmological models (where RLAGN feedback is `sub-grid physics') and
detailed modelling of RLAGN physics. In principle cosmological models
can provide estimates of many of the key quantities (local
environment, time-dependent mass accretion rates, black hole mass and
(vector) spin...) to allow the self-consistent simulation of an entire
mock RLAGN population while also reproducing standard constraints such
as the galaxy luminosity function. This sort of large-volume modelling
work will be key to the interpretation of the very large datasets to
be provided by the next-generation radio and optical surveys.

\section*{Acknowledgements}

We thank an anonymous referee for constructive comments on the
original draft of this review. MJH and JHC acknowledge support from the UK Science and Technology Facilities
Council [grants ST/R000905/1 and ST/R000794/1, respectively]. This work made use of {\sc APLpy}, an open-source plotting package for Python \citep{Robitaille+Bressert12} and the Veusz plotting package written by Jeremy Sanders (\url{https://veusz.github.io/}). We are
grateful to Miranda Jarvis and Tao An for supplying data and code for
the plot shown in Fig.\ \ref{fig:jarvis_plot}.

\bibliographystyle{elsarticle-harv} 
\bibliography{mjh,cards}

\end{document}